\documentclass[preprint,dvipdfmx]{ptephy}

\preprintnumber{KYUSHU-HET-188}

\allowdisplaybreaks[1]

\usepackage{amssymb}
\usepackage{amsthm}
\usepackage{amsmath}
\usepackage{booktabs}
\usepackage{bbm}
\usepackage{mathtools}
\usepackage{subcaption}
\usepackage{braket}
\DeclareMathOperator{\tr}{tr}

\newcommand{\Slash}[1]{{\ooalign{\hfil/\hfil\crcr$#1$}}}
\numberwithin{equation}{section}

\usepackage{url}

\begin{document}

\title{Gradient flow representation of the four-dimensional $\mathcal{N}=2$
super Yang--Mills supercurrent}

\author{%
\name{\fname{Aya} \surname{Kasai}}{1},
\name{\fname{Okuto} \surname{Morikawa}}{1},
and
\name{\fname{Hiroshi} \surname{Suzuki}}{1,\ast}
}

\address{%
\affil{1}{Department of Physics, Kyushu University
744 Motooka, Nishi-ku, Fukuoka, 819-0395, Japan}
\email{hsuzuki@phys.kyushu-u.ac.jp}
}

\date{\today}

\begin{abstract}
In K.~Hieda, A.~Kasai, H.~Makino, and H.~Suzuki, Prog.\ Theor.\ Exp.\ Phys.\
\textbf{2017}, 063B03 (2017), a properly normalized supercurrent in the
four-dimensional (4D) $\mathcal{N}=1$ super Yang--Mills theory (SYM) that works
within on-mass-shell correlation functions of gauge-invariant operators is
expressed in a regularization-independent manner by employing the gradient
flow. In the present paper, this construction is extended to the supercurrent
in the 4D $\mathcal{N}=2$ SYM. The so-constructed supercurrent will be useful,
for instance, for fine tuning of lattice parameters toward the supersymmetric
continuum limit in future lattice simulations of the 4D $\mathcal{N}=2$ SYM.
\end{abstract}

\subjectindex{B01, B16, B31, B32}
\maketitle

\section{Introduction and summary}
\label{sec:1}
In quantum field theory, lattice regularization enables nonperturbative
computation from first principles but it breaks preferred symmetries in the
target theory (such as the chiral and spacetime symmetries) quite often. For
this reason, nonperturbative computation in supersymmetric field theory from
first principles, especially on the basis of the
lattice~\cite{Catterall:2009it,Kadoh:2016eju} is generally difficult, requiring
intricate fine tuning of lattice parameters toward the supersymmetric continuum
limit.

For the supersymmetric continuum limit, the lattice parameters should be tuned
so that the Ward--Takahashi (WT) relations associated with the supersymmetry
(SUSY) hold up to the finite lattice spacing effect. More specifically, one
imposes the conservation law of the supercurrent---the Noether current
associated with SUSY. See Refs.~\cite{Curci:1986sm,Kaplan:1983sk,%
Farchioni:2001wx,Suzuki:2012pc} for theoretical studies on this issue
and~Ref.~\cite{Ali:2018fbq} for a recent actual numerical study in the
four-dimensional (4D) $\mathcal{N}=1$ super Yang--Mills (SYM) theory. To carry
out this program, however, one has to determine not only lattice parameters but
also the supercurrent at the same time because the expression of the Noether
current can be nontrivial when the regularization breaks the associated
symmetry. This fact further complicates the situation.

Having the above problem in mind, in~Ref.~\cite{Hieda:2017sqq} the authors
constructed a regularization-independent expression of the supercurrent in the
4D $\mathcal{N}=1$ SYM by employing the gradient flow~\cite{Narayanan:2006rf,%
Luscher:2009eq,Luscher:2010iy,Luscher:2011bx,Luscher:2013cpa,Luscher:2013vga,%
Ramos:2015dla}. The idea is that since composite operators of fields evolved by
the flow automatically become finite renormalized
operators~\cite{Luscher:2011bx,Luscher:2013cpa} (see
also~Ref.~\cite{Hieda:2016xpq}), the expression of the supercurrent in terms of
flowed fields is independent of regularization (in the limit in which the UV
cutoff is removed); thus the expression is universal. In this way, one can
have, a priori, an expression that becomes \emph{automatically\/} the properly
normalized conserved supercurrent in the continuum limit.

This type of construction of the Noether current by the gradient flow was first
considered for the energy--momentum tensor---the Noether current associated
with the translational invariance~\cite{Suzuki:2013gza,Makino:2014taa}; see
also Ref.~\cite{DelDebbio:2013zaa}. Although in conventional lattice gauge
theory fine tuning for the restoration of the translational invariance is not
necessary because of lattice symmetries, the construction of the associated
Noether current, i.e., the energy--momentum tensor, is still intricate due to
the lack of the corresponding spacetime
symmetry~\cite{Caracciolo:1989pt,Suzuki:2016ytc}. The energy--momentum tensor
carries important physical information and the construction
in~Refs.~\cite{Suzuki:2013gza,Makino:2014taa} has been applied to lattice
simulations in~Refs.~\cite{Asakawa:2013laa,Taniguchi:2016ofw,%
Kitazawa:2016dsl,Ejiri:2017wgd,Kitazawa:2017qab,Kanaya:2017cpp,%
Taniguchi:2017ibr,Yanagihara:2018qqg,Hirakida:2018uoy}; see
also~Ref.~\cite{Morikawa:2018fek}.\footnote{A similar construction has also
been considered for fermion bilinear operators~\cite{Endo:2015iea,%
Hieda:2016lly}, including the (axial) vector current and (pseudo-) scalar
density~\cite{Bochicchio:1985xa}, and has also been applied in lattice
simulations in~Refs.~\cite{Taniguchi:2016ofw,Taniguchi:2016tjc}.}

In this paper, as a natural extension of the study
of~Ref.~\cite{Hieda:2017sqq}, the construction of the supercurrent in the 4D
$\mathcal{N}=2$ SYM~\cite{Ferrara:1974pu,Fayet:1975yi,Brink:1976bc} is
considered.\footnote{This is a natural extension in the sense that we need the
asymptotic freedom for the construction; in the 4D $\mathcal{N}=2$ SYM, all the
interactions are governed by the asymptotic-free gauge coupling.} For this
theory, a highly nontrivial consistent low-energy description has been
known~\cite{Seiberg:1994rs}. Thus it is of great interest to investigate its
low-energy physics by complementary nonperturbative techniques such as the
lattice. See Refs.~\cite{Sugino:2003yb,Sugino:2004uv,Damgaard:2007be,%
Damgaard:2007xi,Hanada:2011qx,Takimi:2012zw} for lattice formulations which are
designed to simplify the fine tuning.

Since our analysis in this paper is rather lengthy, we summarize the basic
line of reasoning and the main results in this section.

Our strategy is basically identical to that of~Ref.~\cite{Hieda:2017sqq}:
First, we need the expression of the properly normalized conserved supercurrent
with a certain regularization. As noted in~Ref.~\cite{Hieda:2017sqq}, this is
already an intricate problem because there is no known regularization that
manifestly preserves the gauge symmetry and SUSY. We adopt the dimensional
regularization for computational ease. Also, since we take the Wess--Zumino
(WZ) gauge~\cite{Wess:1992cp}, having the application to actual lattice
simulations in mind, this gauge choice and the associated gauge fixing and
ghost--anti-ghost terms also break SUSY. The would-be SUSY WT relations are
thus full of SUSY-breaking terms. Only after adding appropriate counterterms to
the action and appropriate rearrangements of terms under the renormalization,
the SUSY emerges in the WT relations. Although this realization of SUSY in
renormalized theory should occur from the general argument
(see~Ref.~\cite{Capri:2018lru} and references cited therein), as demonstrated
in~Ref.~\cite{Hieda:2017sqq} for the 4D $\mathcal{N}=1$ SYM to the one-loop
order,\footnote{This is the first explicit demonstration to our knowledge.} and
as we will see in this paper through~Sects.~\ref{sec:2} to~\ref{sec:6} for
the 4D $\mathcal{N}=2$ SYM to the one-loop order, the realization of SUSY after
the renormalization appears miraculous. Although our argument is parallel to
that of~Ref.~\cite{Hieda:2017sqq}, the required computational labor is much
higher because of the presence of the scalar field. The required Feynman
diagrams for the operator renormalization are collected
in~Appendix~\ref{sec:C}.

In~Sect.~\ref{sec:7}, from the results obtained to that point we determine the
expression of the properly normalized conserved supercurrent that works within
on-mass-shell correlation functions of gauge-invariant operators to the
one-loop order. In~Sect.~\ref{sec:8}, we introduce the gradient flow and the
small flow time expansion~\cite{Luscher:2011bx}, the expansion with respect
to the flow time~$t$. For the 4D $\mathcal{N}=2$ SYM, in addition to the
flow of the gauge and fermion fields frequently considered in the literature,
we have to include the flow of the scalar field. For this, we use a simple flow
equation following the discussion of~Ref.~\cite{Capponi:2015ucc}. See also
Refs.~\cite{Fujikawa:2016qis,Aoki:2016ohw,Makino:2018rys} for related
arguments. We then compute the wave function renormalization of the flowed
fields; although we can almost borrow the result of~Ref.~\cite{Makino:2014taa},
the contribution of the scalar field has to be computed anew as summarized
in~Table~\ref{table:1}.\footnote{Flow equations in supersymmetric theories that
are alternative to our choice are given
in~Refs.~\cite{Kikuchi:2014rla,Aoki:2017iwi}. There is interesting indication
that no wave function renormalization is necessary~\cite{Kadoh:2018} if one
employs the flow equation of~Ref.~\cite{Kikuchi:2014rla}.} We then compute the
small flow time expansion~\cite{Luscher:2011bx} of composite operators relevant
to the representation of the supercurrent. The computation of the small flow
time expansion has been presented many times in above references (see
also~Ref.~\cite{Suzuki:2015bqa}), but the presence of the scalar field greatly
increases the number of required flow Feynman diagrams; the diagrams are
summarized in~Appendix~\ref{sec:C}.

By substituting the small time expansion obtained in~Sect.~\ref{sec:8} in the
supercurrent from~Sect.~\ref{sec:7}, we have the representation of the
supercurrent in terms of the flowed fields. The uncomputed higher-order $O(t)$
terms in the small flow time expansion are neglected by taking the
limit~$t\to0$. At the same time, in the representation of the supercurrent, a
renormalization group argument shows that one can use the running gauge
coupling~$\Bar{g}(\mu)$ in which the renormalization scale~$\mu$ is identified
with~$1/\sqrt{8t}$. In the $t\to0$ limit, therefore, $\Bar{g}(\mu)\to0$ because
of the asymptotic freedom and this justifies the perturbative computation
assumed so far.

In this way, we have the supercurrent
\begin{align}
   &\Tilde{S}_\mu^{\text{imp}}(x)
\notag\\
   &=\lim_{t\to0}\biggl(\left\{
   1+\frac{\Bar{g}^2}{(4\pi)^2}C_2(G)
   \left[-\ln\pi-\frac{9}{4}+\frac{1}{2}\ln(432)\right]
   \right\}\left(-\frac{1}{4\Bar{g}}\right)
   \sigma_{\rho\sigma}\gamma_\mu\mathring{\chi}^aG_{\rho\sigma}^a
\notag\\
   &\qquad\qquad{}
   -\frac{\Bar{g}}{(4\pi)^2}C_2(G)
   \gamma_\nu\mathring{\chi}^a
   G_{\nu\mu}^a
\notag\\
   &\qquad\qquad{}
   +\left\{1+\frac{\Bar{g}^2}{(4\pi)^2}C_2(G)
   \left[-\frac{19}{4}+4\ln2+\frac{1}{2}\ln(432)\right]\right\}
\notag\\
   &\qquad\qquad\qquad\qquad{}
   \times\frac{1}{2\sqrt{2}}
   \left(\frac{1}{3}\sigma_{\mu\nu}-\delta_{\mu\nu}\right)
   (P_+D_\nu\mathring{\chi}^a\mathring{\phi}^a
   -P_-D_\nu\mathring{\chi}^a\mathring{\phi}^{\dagger a})
\notag\\
   &\qquad\qquad{}
   -\frac{3}{\sqrt{2}}\frac{\Bar{g}^2}{(4\pi)^2}C_2(G)
   (P_+D_\mu\mathring{\chi}^a\mathring{\phi}^a
   -P_-D_\mu\mathring{\chi}^a\mathring{\phi}^{\dagger a})
\notag\\
   &\qquad\qquad{}
   +\left\{1+\frac{\Bar{g}^2}{(4\pi)^2}C_2(G)
   \left[\frac{1}{2}+4\ln2+\frac{1}{2}\ln(432)\right]\right\}
\notag\\
   &\qquad\qquad\qquad\qquad{}
   \times\left(-\frac{1}{\sqrt{2}}\right)
   \left(\frac{1}{3}\sigma_{\mu\nu}-\delta_{\mu\nu}\right)
   (P_+\mathring{\chi}^aD_\nu\mathring{\phi}^a
   -P_-\mathring{\chi}^aD_\nu\mathring{\phi}^{\dagger a})
\notag\\
   &\qquad\qquad{}
   +\frac{1}{\sqrt{2}}\frac{\Bar{g}^2}{(4\pi)^2}C_2(G)
   \left(\frac{1}{3}\sigma_{\mu\nu}-\delta_{\mu\nu}\right)
   \gamma_5D_\nu\mathring{\chi}^a(\mathring{\phi}^a+\mathring{\phi}^{\dagger a})
\notag\\
   &\qquad\qquad{}
   +\frac{1}{2\sqrt{2}}\frac{\Bar{g}^2}{(4\pi)^2}C_2(G)
   \left(\frac{1}{3}\sigma_{\mu\nu}-\delta_{\mu\nu}\right)
   \gamma_5\mathring{\chi}^aD_\nu(\mathring{\phi}^a+\mathring{\phi}^{\dagger a})
\notag\\
   &\qquad\qquad{}
   -\frac{1}{4}\frac{\Bar{g}^3}{(4\pi)^2}C_2(G)
   f^{abc}\gamma_5\gamma_\mu
   \mathring{\chi}^a\mathring{\phi}^{\dagger b}\mathring{\phi}^c
   \biggr),
\label{eq:(1.1)}
\end{align}
and its Dirac conjugate
\begin{align}
   &\Tilde{\Bar{S}}_\mu^{\text{imp}}(x)
\notag\\
   &=\lim_{t\to0}\biggr(\left\{
   1+\frac{\Bar{g}^2}{(4\pi)^2}C_2(G)
   \left[-\ln\pi-\frac{9}{4}+\frac{1}{2}\ln(432)\right]
   \right\}\left(-\frac{1}{4\Bar{g}}\right)
   \mathring{\Bar{\chi}}^a\gamma_\mu\sigma_{\rho\sigma}G_{\rho\sigma}^a
\notag\\
   &\qquad\qquad{}
   +\frac{\Bar{g}}{(4\pi)^2}C_2(G)
   \mathring{\Bar{\chi}}^a\gamma_\nu
   G_{\nu\mu}^a
\notag\\
   &\qquad\qquad{}
   +\left\{1+\frac{\Bar{g}^2}{(4\pi)^2}C_2(G)
   \left[-\frac{19}{4}+4\ln2+\frac{1}{2}\ln(432)\right]\right\}
\notag\\
   &\qquad\qquad\qquad\qquad{}
   \times\left(-\frac{1}{2\sqrt{2}}\right)
   (D_\nu\mathring{\Bar{\chi}}^aP_+\mathring{\phi}^a
   -D_\nu\mathring{\Bar{\chi}}^aP_-\mathring{\phi}^{\dagger a})
   \left(\frac{1}{3}\sigma_{\nu\mu}-\delta_{\nu\mu}\right)
\notag\\
   &\qquad\qquad{}
   +\frac{3}{\sqrt{2}}\frac{\Bar{g}^2}{(4\pi)^2}C_2(G)
   (D_\mu\mathring{\Bar{\chi}}^aP_+\mathring{\phi}^a
   -D_\mu\mathring{\Bar{\chi}}^aP_-\mathring{\phi}^{\dagger a})
\notag\\
   &\qquad\qquad{}
   +\left\{1+\frac{\Bar{g}^2}{(4\pi)^2}C_2(G)
   \left[\frac{1}{2}+4\ln2+\frac{1}{2}\ln(432)\right]\right\}
\notag\\
   &\qquad\qquad\qquad\qquad{}
   \times\frac{1}{\sqrt{2}}
   (P_+\mathring{\Bar{\chi}}^aD_\nu\mathring{\phi}^a
   -P_-\mathring{\Bar{\chi}}^aD_\nu\mathring{\phi}^{\dagger a})
   \left(\frac{1}{3}\sigma_{\nu\mu}-\delta_{\nu\mu}\right)
\notag\\
   &\qquad\qquad{}
   -\frac{1}{\sqrt{2}}\frac{\Bar{g}^2}{(4\pi)^2}C_2(G)
   D_\nu\mathring{\Bar{\chi}}^a\gamma_5
   (\mathring{\phi}^a+\mathring{\phi}^{\dagger a})
   \left(\frac{1}{3}\sigma_{\nu\mu}-\delta_{\nu\mu}\right)
\notag\\
   &\qquad\qquad{}
   -\frac{1}{2\sqrt{2}}\frac{\Bar{g}^2}{(4\pi)^2}C_2(G)
   \mathring{\Bar{\chi}}^a\gamma_5
   D_\nu(\mathring{\phi}^a+\mathring{\phi}^{\dagger a})
   \left(\frac{1}{3}\sigma_{\nu\mu}-\delta_{\nu\mu}\right)
\notag\\
   &\qquad\qquad{}
   +\frac{1}{4}\frac{\Bar{g}^3}{(4\pi)^2}C_2(G)
   f^{abc}
   \mathring{\Bar{\chi}}^a\gamma_\mu\gamma_5
   \mathring{\phi}^{\dagger b}\mathring{\phi}^c
   \biggr).
\label{eq:(1.2)}
\end{align}
These are our main results in this paper.\footnote{Our notational convention is
summarized in~Appendix~\ref{sec:A}.} In these expressions, $\Bar{g}$ denotes
the running gauge coupling in the minimal subtraction (MS)
scheme~$\Bar{g}(\mu)$ in which the renormalization scale~$\mu$ is set
to~$\mu=1/\sqrt{8t}$. The beta function in
\begin{equation}
   \mu\frac{d\Bar{g}(\mu)}{d\mu}=\beta(\Bar{g}(\mu))
\label{eq:(1.3)}
\end{equation}
is~$\beta(g)=-2g^3C_2(G)/(4\pi)^2$ to all orders in perturbation
theory~\cite{Tarasov:1980au,Avdeev:1981ew,Grisaru:1982zh,Gates:1983nr,%
Seiberg:1988ur}. In the right-hand sides of~Eqs.~\eqref{eq:(1.1)}
and~\eqref{eq:(1.2)}, the fields~$\chi$, $\Bar{\chi}$, $B_\mu$ ($G_{\mu\nu}$
is the field strength corresponding to~$B_\mu$), $\phi$, and~$\phi^\dagger$ are
fields evolved from the corresponding fields in the original 4D $\mathcal{N}=2$
SYM, $\psi$, $\Bar{\psi}$, $A_\mu$, $\varphi$, and~$\varphi^\dagger$ by flow
equations Eqs.~\eqref{eq:(8.1)}--\eqref{eq:(8.7)} to the flow time~$t$. The
``ring''~$\mathring{\phantom{X}}$ above a field implies that the normalization
of the field is changed as in~Eqs.~\eqref{eq:(8.14)}--\eqref{eq:(8.19)}; this
prescription avoids the explicit wave function
renormalization~\cite{Makino:2014taa}. In this way, the expressions
in~Eqs.~\eqref{eq:(1.1)} and~\eqref{eq:(1.2)} are manifestly finite and
are independent of regularization. In particular, Eqs.~\eqref{eq:(1.1)}
and~\eqref{eq:(1.2)} can be used with the lattice regularization; we believe
that one can use the representation for the fine tuning and/or for extracting
some low-energy physics associated with the supercurrent.

In~Appendix~\ref{sec:B}, we summarize the parity and charge conjugation
symmetries that are quite helpful in the actual calculation.

\section{Four-dimensional $\mathcal{N}=2$ SYM in the WZ gauge}
\label{sec:2}
\subsection{The actions and SUSY transformations}
The Euclidean action of the $\mathcal{N}=2$ SYM in the WZ gauge is given
by~$S=\int d^Dx\,\mathcal{L}$,\footnote{Here, we assume the spacetime dimension
is~$D$ to apply the dimensional regularization.} where
\begin{align}
   \mathcal{L}
   &=\frac{1}{4g_0^2}F_{\mu\nu}^aF_{\mu\nu}^a+\Bar{\psi}^a\Slash{D}\psi^a
   +D_\mu\varphi^{\dagger a}D_\mu\varphi^a
   -\frac{1}{2}g_0^2f^{abc}f^{ade}
   \varphi^{\dagger b}\varphi^c\varphi^{\dagger d}\varphi^e
\notag\\
   &\qquad{}
   +\sqrt{2}g_0f^{abc}\Bar{\psi}^a
   \left(P_+\varphi^b-P_-\varphi^{\dagger b}\right)\psi^c.
\label{eq:(2.1)}
\end{align}
In this expression, $\psi$ ($\Bar{\psi}$) is the Dirac fermion field and
$\varphi$ ($\varphi^\dagger$) the complex scalar field; $F_{\mu\nu}^a$ is the
field strength of the gauge field~$A_\mu^a$,
$F_{\mu\nu}^a=\partial_\mu A_\nu^a-\partial_\nu A_\mu^a+f^{abc}A_\mu^bA_\nu^c$. All
fields belong to the adjoint representation. In four dimensions, i.e.\ when
$D=4$, the action~$S$ is invariant under the following SUSY transformations,
\begin{align}
   \delta_\xi A_\mu^a
   &=\frac{1}{2}g_0\left(
   \Bar{\xi}\gamma_\mu\psi^a-\Bar{\psi}^a\gamma_\mu\xi
   \right),
\label{eq:(2.2)}
\\
   \delta_\xi\varphi^a
   &=\frac{1}{\sqrt{2}}\left(
   -\Bar{\xi}P_-\psi^a+\Bar{\psi}^aP_-\xi
   \right),
\label{eq:(2.3)}
\\
   \delta_\xi\varphi^{\dagger a}
   &=\frac{1}{\sqrt{2}}\left(
   \Bar{\xi}P_+\psi^a-\Bar{\psi}^aP_+\xi
   \right),
\label{eq:(2.4)}
\\
   \delta_\xi\psi^a
   &=-\frac{1}{4g_0}\sigma_{\mu\nu}\xi F_{\mu\nu}^a
   -\frac{1}{\sqrt{2}}\gamma_\mu P_+\xi D_\mu\varphi^a
   +\frac{1}{\sqrt{2}}\gamma_\mu P_-\xi D_\mu\varphi^{\dagger a}
   -\frac{1}{2}g_0\gamma_5\xi f^{abc}\varphi^{\dagger b}\varphi^c,
\label{eq:(2.5)}
\\
   \delta_\xi\Bar{\psi}^a
   &=\frac{1}{4g_0}\Bar{\xi}\sigma_{\mu\nu}F_{\mu\nu}^a
   -\frac{1}{\sqrt{2}}\Bar{\xi}\gamma_\mu P_-D_\mu\varphi^a
   +\frac{1}{\sqrt{2}}\Bar{\xi}\gamma_\mu P_+D_\mu\varphi^{\dagger a}
   -\frac{1}{2}g_0\Bar{\xi}\gamma_5f^{abc}\varphi^{\dagger b}\varphi^c.
\label{eq:(2.6)}
\end{align}
The easiest way to derive these formulas is dimensional reduction from the 6D
$\mathcal{N}=1$ SYM that possesses a much simpler
structure~\cite{Brink:1976bc}.

To apply the perturbation theory, we also introduce the gauge-fixing and
ghost--anti-ghost terms by
\begin{align}
   S_{\text{gf}}
   &=\frac{\lambda_0}{2g_0^2}\int d^Dx\,
   \partial_\mu A_\mu^a\partial_\nu A_\nu^a,
\label{eq:(2.7)}
\\
   S_{c\Bar{c}}
   &=-\frac{1}{g_0^2}\int d^Dx\,
   \Bar{c}^a\partial_\mu D_\mu c^a,
\label{eq:(2.8)}
\end{align}
where $\lambda_0$ is the bare gauge-fixing parameter.

\subsection{Supercurrent and SUSY-breaking terms}
To derive SUSY WT relations, we consider the SUSY
transformations in~Eqs.~\eqref{eq:(2.2)}--\eqref{eq:(2.6)} with localized
transformation parameters, $\xi\to\xi(x)$ and~$\Bar{\xi}\to\Bar{\xi}(x)$.
Under these, the $D$-dimensional action~$S$ changes as
\begin{equation}
   \delta_\xi S
   =-\int d^Dx\,
   \left[
   \Bar{\xi}(\partial_\mu S_\mu+X_{\text{Fierz}})
   +(\partial_\mu\Bar{S}_\mu+\Bar{X}_{\text{Fierz}})\xi
   \right]
\label{eq:(2.9)}
\end{equation}
where the supercurrents are given by
\begin{align}
   S_\mu
   &=-\frac{1}{4g_0}\sigma_{\rho\sigma}\gamma_\mu\psi^aF_{\rho\sigma}^a
\notag\\
   &\qquad{}
   +\frac{1}{\sqrt{2}}\gamma_\nu\gamma_\mu P_+\psi^aD_\nu\varphi^a
   -\frac{1}{\sqrt{2}}\gamma_\nu\gamma_\mu P_-\psi^aD_\nu\varphi^{\dagger a}
   +\frac{1}{2}g_0f^{abc}\gamma_5\gamma_\mu\psi^a\varphi^{\dagger b}\varphi^c,
\label{eq:(2.10)}
\\
   \Bar{S}_\mu
   &=-\frac{1}{4g_0}\Bar{\psi}^a\gamma_\mu\sigma_{\rho\sigma}F_{\rho\sigma}^a
\notag\\
   &\qquad{}
   -\frac{1}{\sqrt{2}}\Bar{\psi}^a\gamma_\mu\gamma_\nu P_+D_\nu\varphi^a
   +\frac{1}{\sqrt{2}}\Bar{\psi}^a\gamma_\mu\gamma_\nu P_-D_\nu\varphi^{\dagger a}
   -\frac{1}{2}g_0f^{abc}\Bar{\psi}^a\gamma_\mu\gamma_5
   \varphi^{\dagger b}\varphi^c,
\label{eq:(2.11)}
\end{align}
while the breaking terms (due to the breaking of the Fierz identity in $D\neq4$
dimensions) are
\begin{align}
   X_{\text{Fierz}}
   &=\frac{1}{2}g_0f^{abc}\gamma_\mu\psi^a\Bar{\psi}^b\gamma_\mu\psi^c
   +\frac{1}{2}g_0f^{abc}\gamma_5\psi^a\Bar{\psi}^b\gamma_5\psi^c
   -\frac{1}{2}g_0f^{abc}\psi^a\Bar{\psi}^b\psi^c,
\label{eq:(2.12)}
\\
   \Bar{X}_{\text{Fierz}}
   &=-\frac{1}{2}g_0f^{abc}\Bar{\psi}^a\gamma_\mu\psi^b\Bar{\psi}^c\gamma_\mu
   -\frac{1}{2}g_0f^{abc}\Bar{\psi}^a\gamma_5\psi^b\Bar{\psi}^c\gamma_5
   +\frac{1}{2}g_0f^{abc}\Bar{\psi}^a\psi^b\Bar{\psi}^c.
\label{eq:(2.13)}
\end{align}

The above supercurrents would be regarded as ``canonical'' ones. From the
perspective of the conformal or scale symmetry of the classical
theory, Eq.~\eqref{eq:(2.1)}, for~$D=4$, it would be natural to use
``improved'' supercurrents defined by
\begin{align}
   S_\mu^{\text{imp}}
   &\equiv S_\mu+\frac{\sqrt{2}}{3}\sigma_{\mu\nu}\partial_\nu
   (P_+\psi^a\varphi^a-P_-\psi^a\varphi^{\dagger a}),
\label{eq:(2.14)}
\\
   \Bar{S}_\mu^{\text{imp}}
   &\equiv\Bar{S}_\mu-\frac{\sqrt{2}}{3}\partial_\nu
   (\Bar{\psi}^aP_+\varphi^a-\Bar{\psi}^aP_-\varphi^{\dagger a})\sigma_{\nu\mu}.
\label{eq:(2.15)}
\end{align}
It can be seen that, noting that $\gamma_\mu\sigma_{\rho\sigma}\gamma_\mu=0$
holds for~$D=4$, these are $\gamma$-traceless,
\begin{equation}
   \gamma_\mu S_\mu^{\text{imp}}=\Bar{S}_\mu^{\text{imp}}\gamma_\mu=0,
\label{eq:(2.16)}
\end{equation}
\emph{under equations of motion}. Note that the terms added
in~Eqs.~\eqref{eq:(2.14)} and~\eqref{eq:(2.15)} do not have the total
divergence and thus $\partial_\mu S_\mu=\partial_\mu S_\mu^{\text{imp}}$
and~$\partial_\mu\Bar{S}_\mu=\partial_\mu\Bar{S}_\mu^{\text{imp}}$.

Through the following analyses, however, we find that $S_\mu^{\text{imp}}$
and~$\Bar{S}_\mu^{\text{imp}}$ are \emph{not\/} finite operators and they can be
rendered finite by adding further terms that are proportional to equations of
motion, as
\begin{align}
   \Tilde{S}_\mu^{\text{imp}}
   &\equiv S_\mu^{\text{imp}}
   -\frac{1}{2\sqrt{2}}\gamma_\mu
   (P_-\Slash{D}\psi^a\varphi^a-P_+\Slash{D}\psi^a\varphi^{\dagger a}
   -\sqrt{2}g_0f^{abc}\gamma_5\psi^a\varphi^{\dagger b}\varphi^c),
\label{eq:(2.17)}
\\
   \Tilde{\Bar{S}}_\mu^{\text{imp}}
   &\equiv\Bar{S}_\mu^{\text{imp}}
   +\frac{1}{2\sqrt{2}}
   (\Bar{\psi}^a\overleftarrow{\Slash{D}}P_-\varphi^a
   -\Bar{\psi}^a\overleftarrow{\Slash{D}}P_+\varphi^{\dagger a}
   -\sqrt{2}g_0f^{abc}\Bar{\psi}^a\gamma_5\varphi^{\dagger b}\varphi^c)
   \gamma_\mu.
\label{eq:(2.18)}
\end{align}
Here, the added term in~Eq.~\eqref{eq:(2.17)} has the structure
\begin{equation}
   (\text{$\varphi$ or~$\varphi^\dagger$})
   \times(\text{the equation of motion of~$\psi$}).
\label{eq:(2.19)}
\end{equation}
The effect of the insertion of such a term in a correlation function can be
deduced by the infinitesimal change of variable~$\Bar{\psi}$ in the functional
integral (i.e., the Schwinger--Dyson equation), where the variation is
proportional to~$\varphi$ or~$\varphi^\dagger$; the associated Jacobian is
unity because $\Bar{\psi}$ and~$\varphi$ (or~$\varphi^\dagger$) are independent
integration variables. Then, \emph{if} the wave function renormalization
factors for~$\Bar{\psi}$ and~$\varphi$ (or~$\varphi^\dagger$) are the same,
then the Schwinger--Dyson equation will show that the
combination in~Eq.~\eqref{eq:(2.19)} is a finite operator. The fact is that the
wave function renormalization factors differ as Eqs.~\eqref{eq:(5.6)}
and~\eqref{eq:(5.7)} show, and the terms added in~Eq.~\eqref{eq:(2.17)} is
diverging and cancels divergences in~$S_\mu^{\text{imp}}$. A similar remark
applies to the added term in~Eq.~\eqref{eq:(2.18)}.

Some calculation shows that these finite supercurrents enjoy extremely simple
forms:
\begin{align}
   \Tilde{S}_\mu^{\text{imp}}
   &=-\frac{1}{4g_0}\sigma_{\rho\sigma}\gamma_\mu\psi^aF_{\rho\sigma}^a
\notag\\
   &\qquad{}
   +\frac{1}{2\sqrt{2}}\left(\frac{1}{3}\sigma_{\mu\nu}-\delta_{\mu\nu}\right)
   (P_+D_\nu\psi^a\varphi^a-P_-D_\nu\psi^a\varphi^{\dagger a})
\notag\\
   &\qquad{}
   -\frac{1}{\sqrt{2}}\left(\frac{1}{3}\sigma_{\mu\nu}-\delta_{\mu\nu}\right)
   (P_+\psi^aD_\nu\varphi^a-P_-\psi^aD_\nu\varphi^{\dagger a}),
\label{eq:(2.20)}
\\
   \Tilde{\Bar{S}}_\mu^{\text{imp}}
   &=-\frac{1}{4g_0}\Bar{\psi}^a\gamma_\mu\sigma_{\rho\sigma}F_{\rho\sigma}^a
\notag\\
   &\qquad{}
   -\frac{1}{2\sqrt{2}}
   (D_\nu\Bar{\psi}^aP_+\varphi^a-D_\nu\Bar{\psi}^aP_-\varphi^{\dagger a})
   \left(\frac{1}{3}\sigma_{\nu\mu}-\delta_{\nu\mu}\right)
\notag\\
   &\qquad{}
   +\frac{1}{\sqrt{2}}
   (\Bar{\psi}^aP_+D_\nu\varphi^a-\Bar{\psi}^aP_-D_\nu\varphi^{\dagger a})
   \left(\frac{1}{3}\sigma_{\nu\mu}-\delta_{\nu\mu}\right).
\label{eq:(2.21)}
\end{align}
At $D=4$, these currents are manifestly $\gamma$-traceless \emph{without any
use of the equation of motion\/} because
$\gamma_\mu\sigma_{\rho\sigma}\gamma_\mu=0$
and~$\gamma_\mu[(1/3)\sigma_{\mu\nu}-\delta_{\mu\nu}]=0$ for~$D=4$.

The gauge-fixing and ghost--anti-ghost terms also break SUSY. We define the
ghost and anti-ghost fields as SUSY singlets. Then
\begin{align}
   \delta_\xi S_{\text{gf}}
   &=-\int d^Dx\,(\Bar{\xi}X_{\text{gf}}+\Bar{X}_{\text{gf}}\xi),
\label{eq:(2.22)}
\\
   \delta_\xi S_{c\Bar{c}}
   &=-\int d^Dx\,(\Bar{\xi}X_{c\Bar{c}}+\Bar{X}_{c\Bar{c}}\xi),
\label{eq:(2.23)}
\end{align}
where
\begin{align}
   X_{\text{gf}}
   &=\frac{\lambda_0}{2g_0}\gamma_\mu\psi^a\partial_\mu\partial_\nu A_\nu^a,
\label{eq:(2.24)}
\\
   \Bar{X}_{\text{gf}}
   &=-\frac{\lambda_0}{2g_0}\Bar{\psi}^a\gamma_\mu
   \partial_\mu\partial_\nu A_\nu^a,
\label{eq:(2.25)}
\end{align}
and
\begin{align}
   X_{c\Bar{c}}
   &=\frac{1}{2g_0}f^{abc}\partial_\mu\Bar{c}^ac^b\gamma_\mu\psi^c,
\label{eq:(2.26)}
\\
   \Bar{X}_{c\Bar{c}}
   &=-\frac{1}{2g_0}f^{abc}\partial_\mu\Bar{c}^ac^b\Bar{\psi}^c\gamma_\mu.
\label{eq:(2.27)}
\end{align}
We note that $X_{\text{gf}}+X_{c\Bar{c}}$ (and also
$\Bar{X}_{\text{gf}}+\Bar{X}_{c\Bar{c}}$) are BRS exact:
\begin{equation}
   X_{\text{gf}}+X_{c\Bar{c}}
   =\delta_B\frac{1}{2g_0}\partial_\mu\Bar{c}^a\gamma_\mu\psi^a,
\label{eq:(2.28)}
\end{equation}
where the BRS transformation~$\delta_B$ is defined by
\begin{align}
   \delta_B A_\mu^a&=D_\mu c^a,&
   \delta_B c^a&=-\frac{1}{2}f^{abc}c^bc^c,
\label{eq:(2.29)}
\\
   \delta_B\Bar{c}^a&=\lambda_0\partial_\mu A_\mu^a,&&
\label{eq:(2.30)}
\\
   \delta_B\psi^a&=-f^{abc}c^b\psi^c,&
   \delta_B\Bar{\psi}^a&=-f^{abc}c^b\Bar{\psi}^c.
\label{eq:(2.31)}
\\
   \delta_B\varphi^a&=-f^{abc}c^b\varphi^c,&
   \delta_B\varphi^{\dagger a}&=-f^{abc}c^b\varphi^{\dagger c}.
\label{eq:(2.32)}
\end{align}
Because of~Eq.~\eqref{eq:(2.28)}, $X_{\text{gf}}+X_{c\Bar{c}}$ does not contribute
in correlation functions of gauge-invariant operators.

\section{SUSY WT relations}
\label{sec:3}
\subsection{SUSY WT relations in bare quantities}
In what follows, we consider SUSY WT relations following from the identities
\begin{equation}
   \left\langle
   \delta_\xi\left[
   \begin{Bmatrix}
   A_\nu^b(y)\\
   \varphi^b(y)\\
   \varphi^{\dagger b}(y)\\
   \end{Bmatrix}
   \Bar{\psi}^c(z)\right]
   \right\rangle
   =0,
\label{eq:(3.1)}
\end{equation}
\begin{equation}
   \left\langle
   \delta_\xi
   \left[
   \Bar{\psi}^b(y)c^c(z)\Bar{c}^d(w)
   \right]
   \right\rangle
   =0.
\label{eq:(3.2)}
\end{equation}
In these identities, the parameters of the SUSY transformation, $\xi$
and~$\Bar{\xi}$, are promoted to local functions, $\xi(x)$
and~$\Bar{\xi}(x)$. The variation of the action~$S+S_{\text{gf}}+S_{c\Bar{c}}$
produces the combination $\partial_\mu S_\mu^{\text{imp}}(x)
+X_{\text{Fierz}}(x)+X_{\text{gf}}(x)+X_{c\Bar{c}}(x)$ as the coefficient
of~$\Bar{\xi}(x)$; recall Eqs.~\eqref{eq:(2.9)}, \eqref{eq:(2.22)},
and~\eqref{eq:(2.23)}. The difference between $\Tilde{S}_\mu^{\text{imp}}(x)$
and $S_\mu^{\text{imp}}(x)$ in~Eq.~\eqref{eq:(2.17)} produces another contact
term that can be determined by considering the variations,
$\delta\Bar{\psi}^a(x)=-1/(2\sqrt{2})\varepsilon(x)\gamma_\mu
[P_-\varphi^a(x)-P_+\varphi^{\dagger a}(x)]$ and $\delta(\text{other fields})=0$,
because the difference is proportional to the equation of motion. In this way,
we have
\begin{align}
   &\left\langle
   \left[
   \partial_\mu\Tilde{S}_\mu^{\text{imp}}(x)
   +X_{\text{Fierz}}(x)+X_{\text{gf}}(x)+X_{c\Bar{c}}(x)
   \right]
   A_\nu^b(y)
   \Bar{\psi}^c(z)
   \right\rangle
\notag\\
   &=-\delta(x-y)\frac{1}{2}g_0
   \left\langle
   \gamma_\nu\psi^b(y)\Bar{\psi}^c(z)
   \right\rangle
\notag\\
   &\qquad{}
   -\delta(x-z)\frac{1}{4g_0}
   \left\langle
   A_\nu^b(y)\sigma_{\rho\sigma}F_{\rho\sigma}^c(z)
   \right\rangle
   +\delta(x-z)\frac{1}{2}g_0
   \left\langle
   A_\nu^b(y)\gamma_5f^{cde}\varphi^{\dagger d}(z)\varphi^e(z)
   \right\rangle
\notag\\
   &\qquad{}
   +\delta(x-z)\frac{1}{\sqrt{2}}
   \left\langle
   A_\nu^b(y)\gamma_\rho
   \left[P_-D_\rho\varphi^c(z)-P_+D_\rho\varphi^{\dagger c}(z)\right]
   \right\rangle
\notag\\
   &\qquad{}
   -\partial_\mu^x\delta(x-z)\frac{1}{2\sqrt{2}}
   \left\langle
   A_\nu^b(y)
   \gamma_\mu
   \left[
   P_-\varphi^c(z)-P_+\varphi^{\dagger c}(z)
   \right]
   \right\rangle,
\label{eq:(3.3)}
\end{align}
and
\begin{align}
   &\left\langle
   \left[
   \partial_\mu\Tilde{S}_\mu^{\text{imp}}(x)
   +X_{\text{Fierz}}(x)+X_{\text{gf}}(x)+X_{c\Bar{c}}(x)
   \right]
   \varphi^b(y)
   \Bar{\psi}^c(z)
   \right\rangle
\notag\\
   &=\delta(x-y)\frac{1}{\sqrt{2}}
   \left\langle
   P_-\psi^b(y)\Bar{\psi}^c(z)
   \right\rangle
\notag\\
   &\qquad{}
   -\delta(x-z)\frac{1}{4g_0}
   \left\langle
   \varphi^b(y)\sigma_{\rho\sigma}F_{\rho\sigma}^c(z)
   \right\rangle
   +\delta(x-z)\frac{1}{2}g_0
   \left\langle
   \varphi^b(y)\gamma_5f^{cde}\varphi^{\dagger d}(z)\varphi^e(z)
   \right\rangle
\notag\\
   &\qquad{}
   +\delta(x-z)\frac{1}{\sqrt{2}}
   \left\langle
   \varphi^b(y)\gamma_\rho
   \left[P_-D_\rho\varphi^c(z)-P_+D_\rho\varphi^{\dagger c}(z)\right]
   \right\rangle
\notag\\
   &\qquad{}
   -\partial_\mu^x\delta(x-z)\frac{1}{2\sqrt{2}}
   \left\langle
   \varphi^b(y)
   \gamma_\mu
   \left[
   P_-\varphi^c(z)-P_+\varphi^{\dagger c}(z)
   \right]
   \right\rangle,
\label{eq:(3.4)}
\end{align}
and
\begin{align}
   &\left\langle
   \left[
   \partial_\mu\Tilde{S}_\mu^{\text{imp}}(x)
   +X_{\text{Fierz}}(x)+X_{\text{gf}}(x)+X_{c\Bar{c}}(x)
   \right]
   \varphi^{\dagger b}(y)
   \Bar{\psi}^c(z)
   \right\rangle
\notag\\
   &=-\delta(x-y)\frac{1}{\sqrt{2}}
   \left\langle
   P_+\psi^b(y)\Bar{\psi}^c(z)
   \right\rangle
\notag\\
   &\qquad{}
   -\delta(x-z)\frac{1}{4g_0}
   \left\langle
   \varphi^{\dagger b}(y)\sigma_{\rho\sigma}F_{\rho\sigma}^c(z)
   \right\rangle
   +\delta(x-z)\frac{1}{2}g_0
   \left\langle
   \varphi^{\dagger b}(y)\gamma_5f^{cde}\varphi^{\dagger d}(z)\varphi^e(z)
   \right\rangle
\notag\\
   &\qquad{}
   +\delta(x-z)\frac{1}{\sqrt{2}}
   \left\langle
   \varphi^{\dagger b}(y)\gamma_\rho
   \left[P_-D_\rho\varphi^c(z)-P_+D_\rho\varphi^{\dagger c}(z)\right]
   \right\rangle
\notag\\
   &\qquad{}
   -\partial_\mu^x\delta(x-z)\frac{1}{2\sqrt{2}}
   \left\langle
   \varphi^{\dagger b}(y)
   \gamma_\mu
   \left[
   P_-\varphi^c(z)-P_+\varphi^{\dagger c}(z)
   \right]
   \right\rangle.
\label{eq:(3.5)}
\end{align}
From Eq.~\eqref{eq:(3.2)}, on the other hand, we have
\begin{align}
   &\left\langle
   \left[
   \partial_\mu\Tilde{S}_\mu^{\text{imp}}(x)
   +X_{\text{Fierz}}(x)+X_{\text{gf}}(x)+X_{c\Bar{c}}(x)
   \right]
   \Bar{\psi}^b(y)c^c(z)\Bar{c}^d(w)
   \right\rangle
\notag\\
   &=-\delta(x-y)\frac{1}{4g_0}
   \left\langle
   \sigma_{\rho\sigma}F_{\rho\sigma}^b(y)
   c^c(z)\Bar{c}^d(w)
   \right\rangle
   +\delta(x-y)\frac{1}{2}g_0
   \left\langle
   \gamma_5f^{bef}\varphi^{\dagger e}(y)\varphi^f(y)
   c^c(z)\Bar{c}^d(w)
   \right\rangle
\notag\\
   &\qquad{}
   +\delta(x-y)\frac{1}{\sqrt{2}}
   \left\langle
   \gamma_\rho\left[P_-D_\rho\varphi^c(z)-P_+D_\rho\varphi^{\dagger c}(z)\right]
   c^c(z)\Bar{c}^d(w)
   \right\rangle
\notag\\
   &\qquad{}
   -\partial_\mu^x\delta(x-y)\frac{1}{2\sqrt{2}}
   \left\langle
   \gamma_\mu
   \left[
   P_-\varphi^b(y)-P_+\varphi^{\dagger b}(y)
   \right]
   c^c(z)\Bar{c}^d(w)
   \right\rangle.
\label{eq:(3.6)}
\end{align}
These are \emph{identities\/} holding exactly under the dimensional
regularization. In what follows, we will rewrite these identities in terms of
renormalized quantities to the one-loop order and find SUSY WT relations among
renormalized quantities. Then, using these SUSY WT relations, we will determine
the form of a properly normalized supercurrent.

\section{The effect of~$X_{\text{Fierz}}$}
\label{sec:4}
\begin{figure}[htbp]
\centering
\begin{subfigure}{0.24\columnwidth}
\centering
\includegraphics[width=0.7\columnwidth]{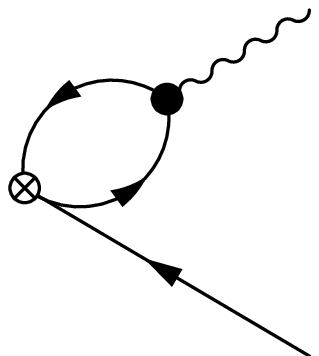}
\end{subfigure}
\hspace*{1em}
\begin{subfigure}{0.24\columnwidth}
\centering
\includegraphics[width=0.7\columnwidth]{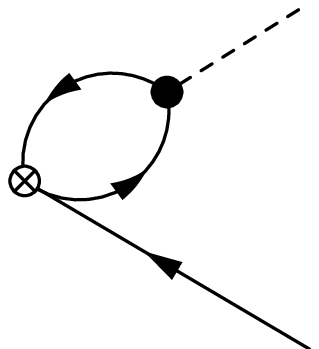}
\end{subfigure}
\caption{One-loop Feynman diagrams containing $X_{\text{Fierz}}$.}
\label{fig:1}
\end{figure}

Before going into the problem of renormalization, we analyze the effect
of~$X_{\text{Fierz}}$ in SUSY WT relations. $X_{\text{Fierz}}$ arises from the
breaking of the Fierz identity at~$D\neq4$ and thus it vanishes in classical
theory at~$D=4$. It survives, however, in quantum theory through UV
divergences. From the calculation of one-loop diagrams in~Fig.~\ref{fig:1} in
the Feynman gauge~$\lambda_0=1$, we have
\begin{align}
   &\left\langle X_{\text{Fierz}}(x)
   \begin{Bmatrix}
   A_\alpha^b(y)\\
   \varphi^b(y)\\
   \varphi^{\dagger b}(y)\\
   \end{Bmatrix}
   \Bar{\psi}^c(z)\right\rangle
\notag\\
   &=\frac{g_0^2}{(4\pi)^2}C_2(G)\delta^{bc}
   \frac{{\mit\Gamma}(D/2)^2}{{\mit\Gamma}(D)}
   {\mit\Gamma}(2-D/2)(-1)(D-4)\int_pe^{ip(x-y)}\int_qe^{iq(x-z)}
\notag\\
   &\qquad{}
   \times\left(\frac{p^2}{4\pi}\right)^{D/2-2}
   \begin{Bmatrix}
   g_0(-p^2\gamma_\alpha+\Slash{p}p_\alpha)\\
   \frac{1}{\sqrt{2}}\frac{1}{D-2}
   \left[(D-1)+3\gamma_5\right]p^2\\
   -\frac{1}{\sqrt{2}}\frac{1}{D-2}
   \left[(D-1)-3\gamma_5\right]p^2\\
   \end{Bmatrix}\frac{1}{p^2}\frac{1}{i\Slash{q}}.
\label{eq:(4.1)}
\end{align}
We see that the effect of~$X_{\text{Fierz}}$ is proportional to~$D-4$ as
expected, but the UV divergences of the diagrams cancel this factor. We can
identify its effect for~$D\to4$ as an insertion of a finite local operator:
\begin{align}
   X_{\text{Fierz}}
   &\stackrel{D\to4}{\to}
   \frac{g_0^2}{(4\pi)^2}C_2(G)
   \left(
   \frac{1}{3g_0}
   \gamma_\nu\psi^a\partial_\mu F_{\mu\nu}^a
   -\frac{1}{\sqrt{2}}
   P_+\psi^a\partial_\mu\partial_\mu\varphi^{\dagger a}
   +\frac{1}{\sqrt{2}}
   P_-\psi^a\partial_\mu\partial_\mu\varphi^a
   \right).
\label{eq:(4.2)}
\end{align}
(Consideration of the correlation function
$\langle X_{\text{Fierz}}(x)\Bar{\psi}^b(y)c^c(z)\Bar{c}^d(w)\rangle$ does not
provide any new information.) We can see that this finite effect
of~$X_{\text{Fierz}}$ in WT relations can be removed by adding a local
counterterm to the original action as~$S\to S+\int d^Dx\,\mathcal{L}'$, where
\begin{align}
   \mathcal{L}'
   &\equiv\frac{g_0^2}{(4\pi)^2}C_2(G)
   \left(
   -\frac{1}{6g_0^2}
   F_{\mu\nu}^aF_{\mu\nu}^a
   +\frac{1}{2}
   \partial_\mu\varphi^a\partial_\mu\varphi^a
   +\frac{1}{2}
   \partial_\mu\varphi^{\dagger a}\partial_\mu\varphi^{\dagger a}
   \right).
\label{eq:(4.3)}
\end{align}
Interestingly, this counterterm breaks the global axial $U(1)$ symmetry,
$\psi\to e^{i\gamma_5\alpha}\psi$, 
$\Bar{\psi}\to\Bar{\psi}e^{i\gamma_5\alpha}$,
$\varphi\to e^{-2i\alpha}\varphi$,
and~$\varphi^\dagger\to e^{2i\alpha}\varphi^\dagger$,
that the original action in~Eq.~\eqref{eq:(2.1)} possesses. The appearance of
such a term in quantum theory is, however, not unexpected because the
dimensional regularization does not preserve this axial $U(1)$ symmetry.

In what follows, we assume the presence of the counterterm~$S'$. This implies
that we can forget about~$X_{\text{Fierz}}$ in WT relations; this is to be
understood throughout the following discussion.

\section{Renormalization}
\label{sec:5}
In this section, we will work out the renormalization in the one-loop order.
We set
\begin{equation}
   D\equiv4-2\epsilon,
\label{eq:(5.1)}
\end{equation}
and assume the MS scheme. We use the abbreviation
\begin{equation}
   \Delta\equiv\frac{g^2}{(4\pi)^2}C_2(G)\frac{1}{\epsilon},
\label{eq:(5.2)}
\end{equation}
where $g$ is the renormalized gauge coupling. Since we work in the one-loop
approximation, we always neglect terms of the order~$O(\Delta^2)$. In
particular, under the product with~$\Delta$, we can always neglect the
difference between bare and renormalized quantities.

The present 4D $\mathcal{N}=2$ SYM is, if the complex scalar field is
neglected, identical to the 4D $\mathcal{N}=1$ SYM except that the fermion
field is Dirac instead of Majorana. So, for diagrams that do not contain
scalar lines, we can borrow the results of~Ref.~\cite{Hieda:2017sqq}, possibly
doubling the fermionic degrees of freedom. This somewhat reduces our labor,
but there still remain many diagrams that contain scalar lines; see
Appendix~\ref{sec:C}. We will work in the Feynman gauge $\lambda_0=1$.

\subsection{Parameters, elementary fields and some composite operators}
For parameters and elementary fields, to the one-loop order, the
renormalization is accomplished by
\begin{align}
   g_0&=\mu^{\epsilon}(1-\Delta)g,
\label{eq:(5.3)}
\\
   \lambda_0&=\lambda,
\label{eq:(5.4)}
\\
   A_\mu^a&=(1-\Delta)A_{\mu R}^a,
\label{eq:(5.5)}
\\
   \begin{Bmatrix}
   \psi^a\\
   \Bar{\psi}^a\\
   \end{Bmatrix}
   &=(1-\Delta)
   \begin{Bmatrix}
   \psi_R^a\\
   \Bar{\psi}_R^a\\
   \end{Bmatrix},
\label{eq:(5.6)}
\\
   \begin{Bmatrix}
   \varphi^a\\
   \varphi^{a\dagger}\\
   \end{Bmatrix}
   &=
   \begin{Bmatrix}
   \varphi_R^a\\
   \varphi_R^{a\dagger}\\
   \end{Bmatrix},
\label{eq:(5.7)}
\\
   c^a&=\left(1-\frac{1}{2}\Delta\right)c_R^a,
\label{eq:(5.8)}
\\
   \Bar{c}^a&=(1-\Delta)\Bar{c}_R^a,
\label{eq:(5.9)}
\end{align}
where the quantities on the right-hand side are renormalized ones.

For some gauge-covariant composite operators that appear on the right-hand
sides of SUSY WT relations, after some calculation, we find
\begin{align}
   F_{\mu\nu}^a
   &=\left(1-\frac{5}{2}\Delta\right)
   (\partial_\mu A_{\nu R}^a-\partial_\nu A_{\mu R}^a)
   +\left(1-\frac{11}{4}\Delta\right)
   \{f^{abc}A_\mu^bA_\nu^c\}_R,
\label{eq:(5.10)}
\\
   f^{abc}\varphi^{\dagger b}\varphi^c
   &=(1-\Delta)\{f^{abc}\varphi^{\dagger b}\varphi^c\}_R,
\label{eq:(5.11)}
\\
   D_\mu\varphi^a
   &=\left(1-\frac{3}{2}\Delta\right)\partial_\mu\varphi_R^a
   +\left(1-\frac{15}{8}\Delta\right)\{f^{abc}A_\mu^b\varphi^c\}_R,
\label{eq:(5.12)}
\\
   D_\mu\varphi^{\dagger a}
   &=\left(1-\frac{3}{2}\Delta\right)\partial_\mu\varphi_R^{\dagger a}
   +\left(1-\frac{15}{8}\Delta\right)\{f^{abc}A_\mu^b\varphi^{\dagger c}\}_R,
\label{eq:(5.13)}
\end{align}
where $\{\mathcal{O}\}_R$ denotes the renormalized composite operator
corresponding to a bare composite operator~$\mathcal{O}$.

\subsection{$X_{\text{gf}}$ and~$X_{c\Bar{c}}$}
From the renormalization factors
in~Eqs.~\eqref{eq:(5.3)}--\eqref{eq:(5.9)} and Eq.~(2.36)
of~Ref.~\cite{Hieda:2017sqq} (divided by~2), and the calculation of the
divergent part of diagrams~A01--A06 in Appendix~\ref{sec:C}, we have the
operator renormalization
\begin{align}
   X_{\text{gf}}+X_{c\Bar{c}}
   &=(1+\Delta)X_{\text{gf}R}+(1-\Delta)X_{c\Bar{c}R}
\notag\\
   &\qquad{}
   +\Delta\partial_\mu
   \left[-\frac{1}{4g}\sigma_{\rho\sigma}\gamma_\mu\psi_R^a
   (\partial_\rho A_{\sigma R}^a-\partial_\sigma A_{\rho R}^a)\right]
\notag\\
   &\qquad{}
   +\Delta\left(-\frac{1}{g}\right)
   \gamma_\nu\psi_R^a\partial_\mu\partial_\mu A_{\nu R}^a
\notag\\
   &\qquad{}
   +\Delta\frac{3}{8g}\sigma_{\mu\nu}\Slash{\partial}\psi_R^a
   (\partial_\mu A_{\nu R}^a-\partial_\nu A_{\mu R}^a)
\notag\\
   &\qquad{}
   +\Delta\frac{1}{8g}\partial_\mu
   \left[
   (A_{\nu R}^a\gamma_\nu\gamma_\mu+2A_{\mu R}^a)\Slash{\partial}\psi_R^a
   \right]
\notag\\
   &\qquad{}
   +\Delta\left(-\frac{1}{\sqrt{2}}\right)\partial_\mu
   \left[
   (P_+\partial_\nu\varphi_R^a-P_-\partial_\nu\varphi_R^{\dagger a})
   \gamma_\nu\gamma_\mu\psi_R^a
   \right]
\notag\\
   &\qquad{}
   +\Delta\left(-\frac{1}{2\sqrt{2}}\right)
   (P_+\varphi_R^a-P_-\varphi_R^{\dagger a})
   \partial_\mu\partial_\mu\psi_R^a
\notag\\
   &\qquad{}
   +\Delta\frac{1}{4}gf^{abc}\gamma_\mu\psi_R^a
   (\varphi_R^{\dagger b}\overleftrightarrow{\partial}_\mu\varphi_R^c)
\notag\\
   &\qquad{}
   +\Delta\left(-\frac{3}{4}\right)gf^{abc}\gamma_5\gamma_\mu\psi_R^a
   \partial_\mu(\varphi_R^{\dagger b}\varphi_R^c)
\notag\\
   &\qquad{}
   +\Delta(-1)gf^{abc}\gamma_5\Slash{\partial}\psi_R^a
   \varphi_R^{\dagger b}\varphi_R^c
\notag\\
   &\qquad{}
   +\Delta\mathcal{H}_1,
\label{eq:(5.14)}
\end{align}
where $X_{\text{gf}R}$ and~$X_{c\Bar{c}R}$ are renormalized finite operators
whose tree-level forms coincide with $X_{\text{gf}}$ and~$X_{c\Bar{c}}$,
respectively. In the last line, $\mathcal{H}_1$ is the abbreviation of possible
``higher order terms,'' which include following (schematically written) types
of operators:
\begin{equation}
   O(\psi_RA_R^2)+O(\psi_RA_R\varphi_R)
   +O(\psi_R\varphi_R^3)+O(\psi_R^3).
\label{eq:(5.15)}
\end{equation}
Our present calculation of diagrams in~Appendix~\ref{sec:C} cannot determine
the coefficients of operators of these forms; they include, for instance,
$\Delta f^{abc}\partial_\mu\psi_RA_{\mu R}^b\varphi^{\dagger c}$,
$\Delta f^{abc}f^{ade}\psi_R^b\varphi_R^c\varphi_R^{\dagger d}\varphi_R^e$,
and~$\Delta f^{abc}\psi_R\Bar{\psi}_R^b\psi_R^c$, etc.

One can confirm that Eq.~\eqref{eq:(5.14)} can be further rewritten in the
following form:
\begin{align}
   &X_{\text{gf}}+X_{c\Bar{c}}
\notag\\
   &=(1+\Delta)
   \left(X_{\text{gf}R}+X_{c\Bar{c}R}\right)
\notag\\
   &\qquad{}
   +\Delta\partial_\mu\Tilde{S}_\mu^{\text{imp}}
   +\Delta\frac{1}{8g}\partial_\mu
   \left[
   (A_{\nu R}^a\gamma_\nu\gamma_\mu+2A_{\mu R}^a)
   \frac{\delta S^t}{\delta\Bar{\psi}^a}
   \right]
\notag\\
   &\qquad{}
   +\Delta g\gamma_\mu\psi_R^a\frac{\delta S^t}{\delta A_\mu^a}
\notag\\
   &\qquad{}
   +\Delta\biggl[
   \frac{3}{8g}
   \sigma_{\mu\nu}(\partial_\mu A_{\nu R}^a-\partial_\nu A_{\mu R}^a)
   -g\gamma_5\{f^{abc}\varphi^{\dagger b}\varphi^c\}_R
\notag\\
   &\qquad\qquad\qquad{}
   -\frac{3}{2\sqrt{2}}
   \gamma_\mu(\partial_\mu\varphi_R^aP_--\partial_\mu\varphi_R^{\dagger a}P_+)
   \biggr]\frac{\delta S^t}{\delta\Bar{\psi}^a}
\notag\\
   &\qquad{}
   +\Delta(-\sqrt{2})P_-\psi_R^a
   \frac{\delta S^t}{\delta\varphi^a}
   +\Delta\sqrt{2}P_+\psi_R^a
   \frac{\delta S^t}{\delta\varphi^{\dagger a}}
\notag\\
   &\qquad{}
   +\Delta\mathcal{H}_2,
\label{eq:(5.16)}
\end{align}
where
\begin{equation}
   S^t\equiv S+S_{\text{gf}}+S_{c\Bar{c}}
\label{eq:(5.17)}
\end{equation}
is the total action.\footnote{In principle, we should include the
counterterm of~Eq.~\eqref{eq:(4.3)} to the total action, but its effect
in~Eq.~\eqref{eq:(5.16)} is~$O(\Delta^2)$ and negligible.} In deriving this
form, we have noted the relation
$\Delta g\gamma_\mu\psi_R^a\delta S_{c\Bar{c}}/(\delta A_\mu^a)=%
-2\Delta X_{c\Bar{c}R}$. The higher-order terms, $\mathcal{H}_2$ in this
expression, which have the same form as~Eq.~\eqref{eq:(5.15)}, differ
from~$\mathcal{H}_1$ in~Eq.~\eqref{eq:(5.14)}.

\subsection{Supercurrent}
Combining the result of~Ref.~\cite{Hieda:2017sqq}, the renormalization factors
in~Eqs.~\eqref{eq:(5.3)}--\eqref{eq:(5.7)} and the calculation of diagrams A01,
A02, B02, B03, B04, C01, C02, and C03 in~Appendix~\ref{sec:C}
(diagrams A03--A06, B16--B19, C06, and~C07 also potentially contribute, but it
turns out that these diagrams do not give divergences), we find
\begin{align}
   \Tilde{S}_\mu^{\text{imp}}
   &=\Tilde{S}_{\mu R}^{\text{imp}}
   +\Delta\frac{1}{4g}\sigma_{\rho\sigma}\gamma_\mu\psi_R^a
   (\partial_\rho A_{\sigma R}^a-\partial_\sigma A_{\rho R}^a)
\notag\\
   &\qquad{}
   +\Delta\frac{3}{4g}
   \left(\frac{1}{3}\sigma_{\mu\sigma}-\delta_{\mu\sigma}\right)
   \gamma_\rho\psi_R^a(\partial_\rho A_{\sigma R}^a-\partial_\nu A_{\sigma R}^a)
\notag\\
   &\qquad{}
   +\Delta\left(-\frac{1}{8g}\right)
   (A_{\nu R}^a\gamma_\nu\gamma_\mu+2A_{\mu R}^a)
   \frac{\delta S^t}{\delta\Bar{\psi}^a}
\notag\\
   &\qquad{}
   +\Delta\mathcal{H}_{3\mu}
\notag\\
   &=\Tilde{S}_{\mu R}^{\text{imp}}
   +\Delta\left(-\frac{1}{8g}\right)
   (A_{\nu R}^a\gamma_\nu\gamma_\mu+2A_{\mu R}^a)
   \frac{\delta S^t}{\delta\Bar{\psi}^a}
   +\Delta\mathcal{H}_{3\mu},
\label{eq:(5.18)}
\end{align}
where $\Tilde{S}_{\mu R}^{\text{imp}}$ is a renormalized composite operator
to the one-loop order and $\mathcal{H}_3$ are again higher-order terms of the
form of~Eq.~\eqref{eq:(5.15)}. The last equality follows from the identity
at~$D=4$:
\begin{equation}
   \left(\frac{1}{3}\sigma_{\mu\sigma}-\delta_{\mu\sigma}\right)\gamma_\rho
   \mathcal{A}_{\rho\sigma}
   =-\frac{1}{3}\sigma_{\rho\sigma}\gamma_\mu\mathcal{A}_{\rho\sigma},
\label{eq:(5.19)}
\end{equation}
where $\mathcal{A}_{\rho\sigma}$ is a quantity that is anti-symmetric
in~$\rho\leftrightarrow\sigma$.

From Eqs.~\eqref{eq:(5.16)} and~\eqref{eq:(5.18)}, we have
\begin{align}
   &\partial_\mu\Tilde{S}_\mu^{\text{imp}}+X_{\text{gf}}+X_{c\Bar{c}}
\notag\\
   &=(1+\Delta)
   \left(\partial_\mu\Tilde{S}_{\mu R}^{\text{imp}}
   +X_{\text{gf}R}+X_{c\Bar{c}R}\right)
\notag\\
   &\qquad{}
   +\Delta g\gamma_\mu\psi_R^a\frac{\delta S^t}{\delta A_\mu^a}
\notag\\
   &\qquad{}
   +\Delta\biggl[
   \frac{3}{8g}
   \sigma_{\mu\nu}(\partial_\mu A_{\nu R}^a-\partial_\nu A_{\mu R}^a)
   -g\gamma_5\{f^{abc}\varphi^{\dagger b}\varphi^c\}_R
\notag\\
   &\qquad\qquad\qquad{}
   -\frac{3}{2\sqrt{2}}
   \gamma_\mu(\partial_\mu\varphi_R^aP_--\partial_\mu\varphi_R^{\dagger a}P_+)
   \biggr]\frac{\delta S^t}{\delta\Bar{\psi}^a}
\notag\\
   &\qquad{}
   +\Delta(-\sqrt{2})P_-\psi_R^a
   \frac{\delta S^t}{\delta\varphi^a}
   +\Delta\sqrt{2}P_+\psi_R^a
   \frac{\delta S^t}{\delta\varphi^{\dagger a}}
\notag\\
   &\qquad{}
   +\Delta\partial_\mu\mathcal{H}_{3\mu}+\Delta\mathcal{H}_2.
\label{eq:(5.20)}
\end{align}
We thus observe that, to the one-loop order, the
combination $\partial_\mu\Tilde{S}_\mu^{\text{imp}}+X_{\text{gf}}+X_{c\Bar{c}}$
appearing in the SUSY WT relations of~Eqs.~\eqref{eq:(3.3)}--\eqref{eq:(3.6)}
(except $X_{\text{Fierz}}$, which was already treated in~Sect.~\ref{sec:4}) is a
linear combination of $\partial_\mu\Tilde{S}_{\mu R}^{\text{imp}}$,
$X_{\text{gf}R}$, and~$X_{c\Bar{c}R}$, up to terms that are proportional to
equations of motion and higher-order terms. This structure is common to the
case of the 4D $\mathcal{N}=1$ SYM and we thus expect that the following
argument will be similar to that of~Ref.~\cite{Hieda:2017sqq}.

\section{SUSY WT relations in renormalized quantities}
\label{sec:6}
We now rewrite the SUSY WT relations of~Eqs.~\eqref{eq:(3.3)}--\eqref{eq:(3.6)}
in terms of renormalized quantities by using Eq.~\eqref{eq:(5.20)} and the
renormalization factors in~Eqs.~\eqref{eq:(5.3)}--\eqref{eq:(5.13)}. Let us
illustrate the calculation in some detail by taking Eq.~\eqref{eq:(3.3)} as the
example: After substituting Eq.~\eqref{eq:(5.20)} in the left-hand side
of~Eq.~\eqref{eq:(3.3)}, we note the Schwinger--Dyson equations:
\begin{align}
   \left\langle
   \mathcal{F}_\mu^a(x)\frac{\delta S^t}{\delta A_\mu^a(x)}
   A_\nu^b(y)\Bar{\psi}^c(z)
   \right\rangle
   &=\delta(x-y)
   \left\langle
   \mathcal{F}_\nu^b(y)\Bar{\psi}^c(z)
   \right\rangle,
\label{eq:(6.1)}
\\
   \left\langle
   \mathcal{F}^a(x)\frac{\delta S^t}{\delta\Bar{\psi}^a(x)}
   \begin{Bmatrix}
   A_\alpha^b(y)\\
   \varphi^b(y)\\
   \varphi^{\dagger b}(y)\\
   \end{Bmatrix}
   \Bar{\psi}^c(z)
   \right\rangle
   &=\delta(x-z)
   \left\langle
   \begin{Bmatrix}
   A_\alpha^b(y)\\
   \varphi^b(y)\\
   \varphi^{\dagger b}(y)\\
   \end{Bmatrix}
   \mathcal{F}^c(z)
   \right\rangle,
\label{eq:(6.2)}
\\
   \left\langle
   \mathcal{F}^a(x)\frac{\delta S^t}{\delta \varphi^a(x)}
   \varphi^b(y)\Bar{\psi}^c(z)
   \right\rangle
   &=\delta(x-y)
   \left\langle
   \mathcal{F}^b(y)\Bar{\psi}^c(z)
   \right\rangle,
\label{eq:(6.3)}
\\
   \left\langle
   \mathcal{F}^a(x)\frac{\delta S^t}{\delta \varphi^{\dagger a}(x)}
   \varphi^{\dagger b}(y)\Bar{\psi}^c(z)
   \right\rangle
   &=\delta(x-y)
   \left\langle
   \mathcal{F}^b(y)\Bar{\psi}^c(z)
   \right\rangle.
\label{eq:(6.4)}
\end{align}
Then, the left-hand side of~Eq.~\eqref{eq:(3.3)} becomes
\begin{align}
   &(1-\Delta)\left\langle
   \left[
   \partial_\mu\Tilde{S}_{\mu R}^{\text{imp}}(x)
   +X_{\text{gf}R}(x)+X_{c\Bar{c}R}(x)
   +\Delta\mathcal{H}_4(x)\right]
   A_{\nu R}^b(y)
   \Bar{\psi}_R^c(z)
   \right\rangle
\notag\\
   &{}
   -2\Delta\delta(x-y)\left(-\frac{1}{2}\right)g
   \left\langle
   \gamma_\nu\psi_R^b(y)\Bar{\psi}_R^c(z)
   \right\rangle
\notag\\
   &{}
   -\frac{3}{2}\Delta\delta(x-z)\left(-\frac{1}{4g}\right)
   \left\langle
   A_{\nu R}^b(y)\sigma_{\rho\sigma}
   \left[\partial_\rho A_{\sigma R}^c(z)-\partial_\sigma A_{\rho R}^c(z)\right]
   \right\rangle
\notag\\
   &{}
   -2\Delta
   \delta(x-z)\frac{1}{2}g
   \left\langle
   A_{\nu R}^b(y)\gamma_5\{f^{cde}\varphi^{\dagger d}\varphi^e\}_R(z)
   \right\rangle
\notag\\
   &{}
   -\frac{3}{2}\Delta\delta(x-z)\frac{1}{\sqrt{2}}
   \left\langle
   A_{\nu R}^b(y)\gamma_\rho
   \left[P_-\partial_\rho\varphi_R^c(z)
   -P_+\partial_\rho\varphi_R^{\dagger c}(z)\right]
   \right\rangle,
\label{eq:(6.5)}
\end{align}
where
\begin{equation}
   \mathcal{H}_4=\partial_\mu\mathcal{H}_{3\mu}+\mathcal{H}_2+O(\Delta).
\label{eq:(6.6)}
\end{equation}

On the other hand, the right-hand side of~Eq.~\eqref{eq:(3.3)} becomes, after
using Eqs.~\eqref{eq:(5.3)}--\eqref{eq:(5.13)},
\begin{align}
   &(1-3\Delta)\delta(x-y)\left(-\frac{1}{2}\right)g
   \left\langle
   \gamma_\nu\psi_R^b(y)\Bar{\psi}_R^c(z)
   \right\rangle
\notag\\
   &{}
   +\left(1-\frac{5}{2}\Delta\right)\delta(x-z)\left(-\frac{1}{4g}\right)
   \left\langle
   A_{\nu R}^b(y)\sigma_{\rho\sigma}
   \left[\partial_\rho A_{\sigma R}^c(z)-\partial_\sigma A_{\rho R}^c(z)
   +\mathcal{H}'(z)\right]
   \right\rangle
\notag\\
   &{}
   +(1-3\Delta)
   \delta(x-z)\frac{1}{2}g
   \left\langle
   A_{\nu R}^b(y)\gamma_5\{f^{cde}\varphi^{\dagger d}\varphi^e\}_R(z)
   \right\rangle
\notag\\
   &{}
   +\left(1-\frac{5}{2}\Delta\right)\delta(x-z)\frac{1}{\sqrt{2}}
   \left\langle
   A_{\nu R}^b(y)\gamma_\rho
   \left[P_-\partial_\rho\varphi_R^c(z)
   -P_+\partial_\rho\varphi_R^{\dagger c}(z)
   +\mathcal{H}'(z)\right]
   \right\rangle
\notag\\
   &{}
   +(1-\Delta)(-1)\partial_\mu^x\delta(x-z)\frac{1}{2\sqrt{2}}
   \left\langle
   A_{\nu R}^b(y)
   \gamma_\mu
   \left[
   P_-\varphi_R^c(z)-P_+\varphi_R^{\dagger c}(z)
   \right]
   \right\rangle,
\label{eq:(6.7)}
\end{align}
where $\mathcal{H}'$ is an abbreviation of ``higher-order terms'' of the
following (schematically written) form,
\begin{equation}
   O(A_R^2)+O(A_R\varphi_R)+O(\varphi_R^3)+O(\psi_R^2).
\label{eq:(6.8)}
\end{equation}
We will be sloppy about the indices of~$\mathcal{H}'$ in order to avoid the
expressions becoming unnecessarily complicated. Then, transposing the last four
lines in~Eq.~\eqref{eq:(6.5)}, the left-hand side of the identity, to the
right-hand side, i.e., Eq.~\eqref{eq:(6.7)}, we find that both sides have
precisely the same overall factor~$1-\Delta$. In this way, we finally have
\begin{align}
   &\left\langle
   \left[
   \partial_\mu\Tilde{S}_{\mu R}^{\text{imp}}(x)
   +X_{\text{gf}R}(x)+X_{c\Bar{c}R}(x)
   +\Delta\mathcal{H}_4(x)\right]
   A_{\nu R}^b(y)
   \Bar{\psi}_R^c(z)
   \right\rangle
\notag\\
   &=\delta(x-y)\left(-\frac{1}{2}\right)g
   \left\langle
   \gamma_\nu\psi_R^b(y)\Bar{\psi}_R^c(z)
   \right\rangle
\notag\\
   &\qquad{}
   +\delta(x-z)\left(-\frac{1}{4g}\right)
   \left\langle
   A_{\nu R}^b(y)\sigma_{\rho\sigma}
   \left[\partial_\rho A_{\sigma R}^c(z)-\partial_\sigma A_{\rho R}^c(z)
   +\mathcal{H}'(z)\right]
   \right\rangle
\notag\\
   &\qquad{}
   +\delta(x-z)\frac{1}{2}g
   \left\langle
   A_{\nu R}^b(y)\gamma_5\{f^{cde}\varphi^{\dagger d}\varphi^e\}_R(z)
   \right\rangle
\notag\\
   &\qquad{}
   +\delta(x-z)\frac{1}{\sqrt{2}}
   \left\langle
   A_{\nu R}^b(y)\gamma_\rho
   \left[P_-\partial_\rho\varphi_R^c(z)
   -P_+\partial_\rho\varphi_R^{\dagger c}(z)
   +\mathcal{H}'(z)\right]
   \right\rangle
\notag\\
   &\qquad{}
   -\partial_\mu^x\delta(x-z)\frac{1}{2\sqrt{2}}
   \left\langle
   A_{\nu R}^b(y)
   \gamma_\mu
   \left[
   P_-\varphi_R^c(z)-P_+\varphi_R^{\dagger c}(z)
   \right]
   \right\rangle.
\label{eq:(6.9)}
\end{align}

Starting from~Eq.~\eqref{eq:(3.4)}, a similar calculation shows that
\begin{align}
   &\left\langle
   \left[
   \partial_\mu\Tilde{S}_{\mu R}^{\text{imp}}(x)
   +X_{\text{gf}R}(x)+X_{c\Bar{c}R}(x)
   +\Delta\mathcal{H}_4(x)\right]
   \varphi_R^b(y)
   \Bar{\psi}_R^c(z)
   \right\rangle
\notag\\
   &=\delta(x-y)\frac{1}{\sqrt{2}}
   \left\langle
   P_-\psi_R^b(y)\Bar{\psi}_R^c(z)
   \right\rangle
\notag\\
   &\qquad{}
   -\delta(x-z)\frac{1}{4g}
   \left\langle
   \varphi_R^b(y)\sigma_{\rho\sigma}
   \left[\partial_\rho A_{\sigma R}^c(z)-\partial_\sigma A_{\rho R}^c(z)
   +\mathcal{H}'(z)\right]
   \right\rangle
\notag\\
   &\qquad{}
   +\delta(x-z)\frac{1}{2}g
   \left\langle
   \varphi_R^b(y)\gamma_5\{f^{cde}\varphi^{\dagger d}\varphi^e\}_R(z)
   \right\rangle
\notag\\
   &\qquad{}
   +\delta(x-z)\frac{1}{\sqrt{2}}
   \left\langle
   \varphi_R^b(y)\gamma_\rho
   \left[P_-\partial_\rho\varphi_R^c(z)
   -P_+\partial_\rho\varphi_R^{\dagger c}(z)
   +\mathcal{H}'(z)\right]
   \right\rangle
\notag\\
   &\qquad{}
   -\partial_\mu^x\delta(x-z)\frac{1}{2\sqrt{2}}
   \left\langle
   \varphi_R^b(y)
   \gamma_\mu
   \left[
   P_-\varphi_R^c(z)-P_+\varphi_R^{\dagger c}(z)
   \right]
   \right\rangle,
\label{eq:(6.10)}
\end{align}
and, for~Eq.~\eqref{eq:(3.5)}, similarly
\begin{align}
   &\left\langle
   \left[
   \partial_\mu\Tilde{S}_{\mu R}^{\text{imp}}(x)
   +X_{\text{gf}R}(x)+X_{c\Bar{c}R}(x)
   +\Delta\mathcal{H}_4(x)\right]
   \varphi_R^{\dagger b}(y)
   \Bar{\psi}_R^c(z)
   \right\rangle
\notag\\
   &=-\delta(x-y)\frac{1}{\sqrt{2}}
   \left\langle
   P_+\psi_R^b(y)\Bar{\psi}_R^c(z)
   \right\rangle
\notag\\
   &\qquad{}
   -\delta(x-z)\frac{1}{4g}
   \left\langle
   \varphi_R^{\dagger b}(y)\sigma_{\rho\sigma}
   \left[\partial_\rho A_{\sigma R}^c(z)-\partial_\sigma A_{\rho R}^c(z)
   +\mathcal{H}'(z)\right]
   \right\rangle
\notag\\
   &\qquad{}
   +\delta(x-z)\frac{1}{2}g
   \left\langle
   \varphi_R^{\dagger b}(y)\gamma_5\{f^{cde}\varphi^{\dagger d}\varphi^e\}_R(z)
   \right\rangle
\notag\\
   &\qquad{}
   +\delta(x-z)\frac{1}{\sqrt{2}}
   \left\langle
   \varphi_R^{\dagger b}(y)\gamma_\rho
   \left[P_-\partial_\rho\varphi_R^c(z)
   -P_+\partial_\rho\varphi_R^{\dagger c}(z)
   +\mathcal{H}'(z)\right]
   \right\rangle
\notag\\
   &\qquad{}
   -\partial_\mu^x\delta(x-z)\frac{1}{2\sqrt{2}}
   \left\langle
   \varphi_R^{\dagger b}(y)
   \gamma_\mu
   \left[
   P_-\varphi_R^c(z)-P_+\varphi_R^{\dagger c}(z)
   \right]
   \right\rangle.
\label{eq:(6.11)}
\end{align}

Also, Eq.~\eqref{eq:(3.6)} yields
\begin{align}
   &\left\langle
   \left[
   \partial_\mu\Tilde{S}_{\mu R}^{\text{imp}}(x)
   +X_{\text{gf}R}(x)+X_{c\Bar{c}R}(x)
   +\Delta\mathcal{H}_4(x)\right]
   \Bar{\psi}_R^b(y)c_R^c(z)\Bar{c}_R^d(w)
   \right\rangle
\notag\\
   &=-\delta(x-y)\frac{1}{4g}
   \left\langle
   \sigma_{\rho\sigma}
   \left[\partial_\rho A_{\sigma R}^b(y)-\partial_\sigma A_{\rho R}^b(y)
   +\mathcal{H}'(y)\right]
   c_R^c(z)\Bar{c}_R^d(w)
   \right\rangle
\notag\\
   &\qquad{}
   +\delta(x-y)\frac{1}{2}g
   \left\langle
   \gamma_5\{f^{bef}\varphi^{\dagger e}\varphi^f\}_R(y)
   c_R^c(z)\Bar{c}_R^d(w)
   \right\rangle
\notag\\
   &\qquad{}
   +\delta(x-y)\frac{1}{\sqrt{2}}
   \left\langle
   \gamma_\rho\left[P_-\partial_\rho\varphi_R^c(y)
   -P_+\partial_\rho\varphi_R^{\dagger c}(y)
   +\mathcal{H}'(y)\right]
   c_R^c(z)\Bar{c}_R^d(w)
   \right\rangle
\notag\\
   &\qquad{}
   -\partial_\mu^x\delta(x-y)\frac{1}{2\sqrt{2}}
   \left\langle
   \gamma_\mu
   \left[
   P_-\varphi_R^b(y)-P_+\varphi_R^{\dagger b}(y)
   \right]
   c_R^c(z)\Bar{c}_R^d(w)
   \right\rangle.
\label{eq:(6.12)}
\end{align}

Thus, in all the above WT relations, we have observed that the combination
\begin{equation}
   \partial_\mu\Tilde{S}_{\mu R}^{\text{imp}}
   +X_{\text{gf}R}+X_{c\Bar{c}R}+\Delta\mathcal{H}_4
\label{eq:(6.13)}
\end{equation}
generates \emph{properly normalized super transformations on renormalized
elementary fields}. The existence of such a finite operator would be expected
on general grounds (i.e., SUSY should be free from the anomaly). Nevertheless,
the validity of renormalized SUSY WT relations in the WZ gauge that we have
observed appears miraculous, because it resulted from nontrivial
renormalization/mixing of various composite operators.

\section{Properly normalized supercurrent}
\label{sec:7}
We have observed that the combination in~Eq.~\eqref{eq:(6.13)} generates the
correct super transformations on renormalized elementary fields. It is by no
means obvious if the combination in~Eq.~\eqref{eq:(6.13)} also generates
correct renormalized SUSY transformations on renormalized \emph{composite
operators}. To answer this, we will need further complicated analyses of SUSY
WT relations containing composite operators. Therefore, as
in~Ref.~\cite{Hieda:2017sqq}, we will be satisfied by finding the form of a
properly normalized supercurrent that works within the on-mass-shell
correlation functions containing gauge-invariant operators. By
``on-mass-shell,'' we mean that all (renormalized) composite operators
including the combination in~Eq.~\eqref{eq:(6.13)} are separated from each
other in position space. In such on-mass-shell correlation functions, we can
still regard the combination in~Eq.~\eqref{eq:(6.13)} as properly normalized
because no UV divergence associated with composite operators colliding at an
equal point arises. In what follows, we show that an insertion of the
combination in~Eq.~\eqref{eq:(6.13)} in such correlation functions reduces to
\begin{equation}
   \partial_\mu\Tilde{S}_\mu^{\text{imp}},
\label{eq:(7.1)}
\end{equation}
where $\Tilde{S}_\mu^{\text{imp}}$ is the supercurrent
in~Eq.~\eqref{eq:(2.20)} (its conjugate is given
by~$\Tilde{\Bar{S}}_\mu^{\text{imp}}$ in~Eq.~\eqref{eq:(2.21)}). This also
implies the conservation law of the current~$\Tilde{S}_\mu^{\text{imp}}$ in
on-mass-shell correlation functions, because for on-mass-shell correlation
functions there will be no contact terms, such as the right-hand side
of~Eq.~\eqref{eq:(6.9)}.

Now, in on-mass-shell correlation functions, equations of motion identically
hold. Under tree-level equations of motion, Eq.~\eqref{eq:(5.16)} then reduces
to
\begin{equation}
   X_{\text{gf}}+X_{c\Bar{c}}
   =X_{\text{gf}R}+X_{c\Bar{c}R}
   +\Delta
   \left(\partial_\mu\Tilde{S}_\mu^{\text{imp}}+X_{\text{gf}}+X_{c\Bar{c}}\right)
   +\Delta\mathcal{H}_2.
\label{eq:(7.2)}
\end{equation}
Moreover, since $\partial_\mu\Tilde{S}_\mu^{\text{imp}}+X_{\text{gf}}+X_{c\Bar{c}}=0$
under tree-level equations of motion, we can further set
\begin{equation}
   X_{\text{gf}}+X_{c\Bar{c}}
   =X_{\text{gf}R}+X_{c\Bar{c}R}+\Delta\mathcal{H}_2
\label{eq:(7.3)}
\end{equation}
in on-mass-shell correlation functions. This, however, identically vanishes in
correlation functions with gauge-invariant operators, because
$X_{\text{gf}}+X_{c\Bar{c}}$ is BRS exact, as noted in~Eq.~\eqref{eq:(2.28)}.
Thus, in on-mass-shell correlation functions with gauge-invariant operators,
the combination in~Eq.~\eqref{eq:(6.13)} can be replaced by
\begin{equation}
   \partial_\mu
   \left(\Tilde{S}_{\mu R}^{\text{imp}}+\Delta\mathcal{H}_{3\mu}\right),
\label{eq:(7.4)}
\end{equation}
where we have used~Eq.~\eqref{eq:(6.6)}. Then, going back
to~Eq.~\eqref{eq:(5.18)}, under tree-level equations of motion, we see that
\begin{equation}
   \Tilde{S}_\mu^{\text{imp}}=\Tilde{S}_{\mu R}^{\text{imp}}
   +\Delta\mathcal{H}_{3\mu}.
\label{eq:(7.5)}
\end{equation}
This is the current appearing in~Eq.~\eqref{eq:(7.4)}, and shows the
above~Eq.~\eqref{eq:(7.1)}.

The bottom line of the above very lengthy one-loop analysis is that, in
on-mass-shell correlation functions that contain gauge-invariant operators
only, the combination in~Eq.~\eqref{eq:(6.13)}, which generates correct
renormalized SUSY transformations on renormalized elementary fields, is
replaced by~$\partial_\mu\Tilde{S}_\mu^{\text{imp}}$. This shows that to the
one-loop order, the bare supercurrent and its conjugate,
\begin{equation}
   \Tilde{S}_\mu^{\text{imp}},\qquad
   \Tilde{\Bar{S}}_\mu^{\text{imp}},
\label{eq:(7.6)}
\end{equation}
can be regarded as the properly normalized supercurrents in on-mass-shell
correlation functions containing only gauge-invariant operators.

We now express these currents by fields defined by the gradient flow.

\section{Gradient flow and the small flow time expansion}
\label{sec:8}
\subsection{Flow equations and the wave function renormalization of flowed
fields}
Our flow equations for the gauge field and the fermion field are
standard ones~\cite{Narayanan:2006rf,Luscher:2009eq,Luscher:2010iy,%
Luscher:2011bx,Luscher:2013cpa}: Let $t\geq0$ be the flow time, for the gauge
field,\footnote{The term that is proportional to the ``gauge-fixing
parameter''~$\alpha_0$ is introduced to simplify the perturbative treatment of
the gauge degrees of freedom. Although this term breaks the gauge covariance,
it can be shown that any gauge-invariant quantity is independent
of~$\alpha_0$~\cite{Luscher:2010iy,Luscher:2011bx}. This gauge-breaking term is
thus physically irrelevant.}
\begin{equation}
   \partial_tB_\mu(t,x)
   =D_\nu G_{\nu\mu}(t,x)+\alpha_0D_\mu\partial_\nu B_\nu(t,x),
   \qquad
   B_\mu(t=0,x)=A_\mu(x),
\label{eq:(8.1)}
\end{equation}
where
\begin{equation}
   D_\mu\equiv\partial_\mu+[B_\mu,\cdot],\qquad
   G_{\mu\nu}(t,x)
   \equiv\partial_\mu B_\nu(t,x)-\partial_\nu B_\mu(t,x)
   +[B_\mu(t,x),B_\nu(t,x)],
\label{eq:(8.2)}
\end{equation}
and, for the fermion fields,
\begin{align}
   \partial_t\chi(t,x)&=D_\mu D_\mu\chi(t,x),&
   \chi(t=0,x)&=\psi(x),
\label{eq:(8.3)}
\\
   \partial_t\Bar{\chi}(t,x)
   &=\Bar{\chi}(t,x)
   \overleftarrow{D}_\mu\overleftarrow{D}_\mu,&
   \Bar{\chi}(t=0,x)&=\Bar{\psi}(x),
\label{eq:(8.4)}
\end{align}
where
\begin{equation}
   D_\mu\equiv\partial_\mu+[B_\mu,\cdot],\qquad
   \overleftarrow{D}_\mu\equiv\overleftarrow{\partial}_\mu-[\cdot,B_\mu].
\label{eq:(8.5)}
\end{equation}
The fields $B_\mu(t,x)$, $\chi(t,x)$, and~$\Bar{\chi}(t,x)$ are referred to as
flowed fields throughout this paper. Although we often call the above
one-parameter evolution the gradient flow for historical reasons, the
right-hand side of these equations is not the equation of motion of the
corresponding field and thus in this sense the evolution is not gradient flow
defined by the functional derivative of the action~$S$. This point does not
matter, however, for the application in the present paper.

For the scalar field, we adopt
\begin{align}
   \partial_t\phi(t,x)&=D_\mu D_\mu\phi(t,x),&
   \phi(t=0,x)&=\varphi(x),
\label{eq:(8.6)}
\\
   \partial_t\phi^\dagger(t,x)
   &=\phi^\dagger(t,x)
   \overleftarrow{D}_\mu\overleftarrow{D}_\mu,&
   \phi^\dagger(t=0,x)&=\varphi^\dagger(x).
\label{eq:(8.7)}
\end{align}
This is also not the gradient flow in the narrow sense. In this case, for the
renormalizability of the flowed scalar field, it is important not to include
other terms of the equation of motion (such as the term arising from the Yukawa
coupling) on the right-hand side of the flow equations. We refer the reader to
Ref.~\cite{Capponi:2015ucc} for the renormalizability of the flow in the scalar
field theory. See also Ref.~\cite{Fujikawa:2016qis,Aoki:2016ohw} for related
studies.

A remarkable feature of the ``gradient'' flow is that any composite operator of
flowed fields for~$t>0$ becomes UV finite (i.e., automatically renormalized)
under the conventional parameter renormalization such as~Eqs.~\eqref{eq:(5.3)}
and~\eqref{eq:(5.4)},\footnote{It is shown that the parameter~$\alpha_0$
in~Eq.~\eqref{eq:(8.1)} does not receive the renormalization.} and wave
function renormalizations of \emph{elementary\/} flowed
fields~\cite{Luscher:2011bx,Luscher:2013cpa}. See
also~Ref.~\cite{Hieda:2016xpq}. Moreover the flowed gauge field does not need
the wave function renormalization~\cite{Luscher:2011bx}. One-loop calculations
in the ``Feynman gauge'' $\lambda_0=\alpha_0=1$ yield
\begin{align}
   \left\langle B_\mu^a(t,x)B_\nu^b(s,y)\right\rangle
   &=g_0^2\delta^{ab}\delta_{\mu\nu}\int_p\,e^{ip(x-y)}\frac{e^{-(t+s)p^2}}{p^2}
   (1+2\Delta)+\text{finite part},
\label{eq:(8.8)}
\\
   \left\langle\chi^a(t,x)\Bar{\chi}^b(s,y)\right\rangle
   &=\delta^{ab}\int_p\,e^{ip(x-y)}\frac{e^{-(t+s)p^2}}{i\Slash{p}}
   (1-4\Delta)+\text{finite part},
\label{eq:(8.9)}
\\
   \left\langle\phi^a(t,x)\phi^{\dagger b}(s,y)\right\rangle
   &=\delta^{ab}\int_p\,e^{ip(x-y)}\frac{e^{-(t+s)p^2}}{p^2}
   (1-2\Delta)+\text{finite part},
\label{eq:(8.10)}
\end{align}
where $\Delta$ is defined by~Eq.~\eqref{eq:(5.2)}. For perturbation calculation
of the correlation function of flowed fields, we refer the reader
to~Refs.~\cite{Luscher:2010iy,Luscher:2011bx,Luscher:2013cpa,Suzuki:2013gza,%
Makino:2014taa}. In~Eq.~\eqref{eq:(8.8)}, we see that the one-loop divergence
is actually removed by the one-loop gauge-coupling renormalization
in~Eq.~\eqref{eq:(5.3)}, and the flowed gauge field does not need the wave
function renormalization. For other fields, setting
\begin{align}
   \chi_R(t,x)&=Z_\chi^{1/2}\chi(t,x),&
   \Bar{\chi}_R(t,x)&=Z_\chi^{1/2}\Bar{\chi}(t,x),
\label{eq:(8.11)}
\\
   \phi_R(t,x)&=Z_\phi^{1/2}\phi(t,x),&
   \phi_R^\dagger(t,x)&=Z_\phi^{1/2}\phi^\dagger(t,x),
\label{eq:(8.12)}
\end{align}
we see, to the one-loop order,
\begin{equation}
   Z_\chi=1+4\Delta,\qquad Z_\phi=1+2\Delta.
\label{eq:(8.13)}
\end{equation}

\subsection{Ringed fields}
\label{sec:8.2}
Although the wave function renormalization of flowed fields renders all
composite operators finite, the wave function renormalization factors
themselves depend on the regularization and this is not satisfactory from the
perspective of a universal representation of composite operators. To avoid
this point, we introduce the following ringed fields,
following~Ref.~\cite{Makino:2014taa}:
\begin{align}
   \mathring{\chi}(t,x)
   &\equiv\sqrt{\frac{-2\dim(G)}
   {(4\pi)^2t^2
   \left\langle\Bar{\chi}^a(t,x)\overleftrightarrow{\Slash{D}}\chi^a(t,x)
   \right\rangle}}
   \,\chi(t,x),
\label{eq:(8.14)}
\\
   \mathring{\Bar{\chi}}(t,x)
   &\equiv\sqrt{\frac{-2\dim(G)}
   {(4\pi)^2t^2
   \left\langle\Bar{\chi}^a(t,x)\overleftrightarrow{\Slash{D}}\chi^a(t,x)
   \right\rangle}}
   \,\Bar{\chi}(t,x).
\label{eq:(8.15)}
\end{align}
The wave function renormalization factor is canceled out in the ringed fields,
and any composite operator of the ringed fields becomes finite without an
explicit wave function renormalization. The expectation value
$\langle\Bar{\chi}(t,x)\overleftrightarrow{\Slash{D}}\chi(t,x)\rangle$ in the
denominator does not vanish. In fact, to the one-loop order (this includes the
contribution of scalar fields in addition to the result
of~Ref.~\cite{Makino:2014taa}),
\begin{align}
   &\left\langle\Bar{\chi}^a(t,x)\overleftrightarrow{\Slash{D}}\chi^a(t,x)
   \right\rangle
\notag\\
   &=\frac{-2\dim(G)}{(4\pi)^2t^2}
   \left\{(8\pi t)^\epsilon
   +\frac{g_0^2}{(4\pi)^2}C_2(G)
   \left[-\frac{4}{\epsilon}-8\ln(8\pi t)-\frac{3}{2}+\ln(432)\right]
   \right\}.
\label{eq:(8.16)}
\end{align}
Similarly, for the scalar fields, we introduce~\cite{Makino:2018rys}
\begin{align}
   \mathring{\phi}(t,x)
   &\equiv\sqrt{\frac{\dim(G)}
   {2(4\pi)^2t
   \left\langle\phi^{\dagger a}(t,x)\phi^a(t,x)
   \right\rangle}}
   \,\phi(t,x),
\label{eq:(8.17)}
\\
   \mathring{\phi}^\dagger(t,x)
   &\equiv\sqrt{\frac{\dim(G)}
   {2(4\pi)^2t
   \left\langle\phi^{\dagger a}(t,x)\phi^a(t,x)
   \right\rangle}}
   \,\phi^\dagger(t,x).
\label{eq:(8.18)}
\end{align}
For the following calculations, we need to compute the expectation value
$\langle\phi^{a\dagger}(t,x)\phi^a(t,x)\rangle$. By using the integration
formula, Eq.~(B2) of~Ref.~\cite{Makino:2014taa}, we have the numbers tabulated
in~Table~\ref{table:1}.
\begin{table}
\caption{Contribution of each diagram in Appendix~\ref{sec:C}
to~Eq.~\eqref{eq:(8.19)} in units
of~$\frac{\dim(G)}{2(4\pi)^2t}\frac{g_0^2}{(4\pi)^2}C_2(G)$.}
\label{table:1}
\begin{center}
\renewcommand{\arraystretch}{2.2}
\setlength{\tabcolsep}{20pt}
\begin{tabular}{cr}
\toprule
 Diagram & \\
\midrule
 E02 & $\dfrac{2}{\epsilon}+4\ln(8\pi t)+6$ \\
 E03 & $\dfrac{2}{\epsilon}+4\ln(8\pi t)+6$ \\
 E04 & $-2-4\ln2+6\ln3$ \\
 E05 & $12\ln2-6\ln3$ \\
 E06 & $-\dfrac{4}{\epsilon}-8\ln(8\pi t)-6$ \\
 E07 & $-\dfrac{2}{\epsilon}-4\ln(8\pi t)-7$ \\
\bottomrule
\end{tabular}
\end{center}
\end{table}
In total,
\begin{align}
   &\left\langle\phi^{\dagger a}(t,x)\phi^a(t,x)\right\rangle
\notag\\
   &=\frac{\dim(G)}{2(4\pi)^2t}
   \left\{\frac{1}{1-\epsilon}(8\pi t)^\epsilon
   +\frac{g_0^2}{(4\pi)^2}C_2(G)
   \left[-\frac{2}{\epsilon}-4\ln(8\pi t)-3+8\ln2\right]\right\}.
\label{eq:(8.19)}
\end{align}

\subsection{Small flow time expansion}
In this subsection we present the computation of the small flow time
expansion~\cite{Luscher:2011bx} of composite operators that are relevant to the
construction of the supercurrents in~Eqs.~\eqref{eq:(2.20)}
and~\eqref{eq:(2.21)}. In what follows, we set
\begin{equation}
   \xi(t)\equiv\frac{g_0^2}{(4\pi)^2}C_2(G)(8\pi t)^{2-D/2}.
\label{eq:(8.20)}
\end{equation}

The calculations of the small flow time expansion are presented
in~Refs.~\cite{Luscher:2011bx,Suzuki:2013gza,Makino:2014taa,Endo:2015iea,%
Suzuki:2015bqa,Hieda:2016lly,Hieda:2017sqq}, and we refer the reader to
these references for the actual computation. In particular, the background
field method developed in~Ref.~\cite{Suzuki:2015bqa} is very powerful and was
applied to the computation of the supercurrent in the 4D $\mathcal{N}=1$ SYM
in~Ref.~\cite{Hieda:2017sqq}. In this paper, however, we do not use this method
because the presence of the scalar field reduces the simplicity of the method;
we thus use the standard diagrammatic expansion as
in~Refs.~\cite{Luscher:2011bx,Makino:2014taa}. The diagrams relevant to the
computation of the supercurrent are collected
in~Appendix~\ref{sec:C};\footnote{If only the ``topology'' of the diagram is
concerned, there also exist other diagrams which are not included
in~Appendix~\ref{sec:C}. We carefully confirmed that those omitted diagrams
give only higher-order contribution in the small flow time expansion. Examples
of such diagrams are B13, B14, and~B15, which do not contribute to the
following expansion.} our convention for the flow Feynman diagram is also
summarized at the beginning of~Appendix~\ref{sec:C}.

Now, for diagrams \emph{without\/} external scalar lines, we can literally use
the results of~Eqs.~(3.7), (3.30), (3.31), and~(3.32)
from~Ref.~\cite{Hieda:2017sqq}, because there is no one-loop diagram that
contains a scalar loop and has gauge external lines only. For diagrams with
external scalar lines, diagrams A01--A06 in~Appendix~\ref{sec:C} contribute.
After some calculation, we have
\begin{align}
   &\frac{1}{g_0}\chi^a(t,x)G_{\mu\nu}^a(t,x)
\notag\\
   &=\left[1+\frac{-2}{D-4}\xi(t)\right]
   \frac{1}{g_0}\psi^a(x)F_{\mu\nu}^a(x)
\notag\\
   &\qquad{}
   +\xi(t)
   \biggl\{
   \frac{2}{(D-4)(D-2)}
   \frac{1}{g_0}\left[
   \gamma_\mu\gamma_\rho\psi^a(x)F_{\rho\nu}^a(x)
   -\gamma_\nu\gamma_\rho\psi^a(x)F_{\rho\mu}^a(x)\right]
\notag\\
   &\qquad\qquad\qquad{}
   +\frac{4}{(D-4)(D-2)D}
   \frac{1}{g_0}\sigma_{\rho\sigma}\sigma_{\mu\nu}\psi^a(x)F_{\rho\sigma}^a(x)
   \biggr\}
\notag\\
   &\qquad{}
   +\xi(t)\sqrt{2}
   \biggl\{
   \frac{4}{(D-4)(D-2)D}\gamma_\rho\gamma_\mu\gamma_\nu
   \left[
   P_+\psi^a(x)D_\rho\varphi^a(x)-P_-\psi^a(x)D_\rho\varphi^{\dagger a}(x)
   \right]
\notag\\
   &\qquad\qquad\qquad\qquad{}
   +\frac{-2}{(D-2)D}\gamma_\nu
   \left[
   P_+\psi^a(x)D_\mu\varphi^a(x)-P_-\psi^a(x)D_\mu\varphi^{\dagger a}(x)
   \right]
\notag\\
   &\qquad\qquad\qquad\qquad{}
   +\frac{-2}{(D-4)(D-2)}\gamma_\nu
   \left[
   P_+D_\mu\psi^a(x)\varphi^a(x)-P_-D_\mu\psi^a(x)\varphi^{\dagger a}(x)
   \right]
\notag\\
   &\qquad\qquad\qquad\qquad{}
   +\frac{2(D+4)}{(D-2)D(D+2)}\gamma_\nu\gamma_5D_\mu\psi^a(x)
   \left[
   \varphi^a(x)+\varphi^{\dagger a}(x)
   \right]
\notag\\
   &\qquad\qquad\qquad\qquad{}
   +\frac{2}{(D-2)(D+2)}\gamma_\nu\gamma_5\psi^a(x)
   D_\mu\left[
   \varphi^a(x)+\varphi^{\dagger a}(x)
   \right]
   \biggr\}
   -(\mu\leftrightarrow\nu)
\notag\\
   &\qquad{}
   +\xi(t)
   \frac{8}{(D-4)(D-2)D}
   g_0f^{abc}\sigma_{\mu\nu}\gamma_5\psi^a(x)\varphi^{\dagger b}(x)\varphi^c(x)
   +O(t).
\label{eq:(8.21)}
\end{align}

For $\chi^a(t,x)D_\mu\phi^a(t,x)$, diagrams B01--B20 and C01--C07
in~Appendix~\ref{sec:C} contribute and we have
\begin{align}
   &\chi^a(t,x)D_\mu\phi^a(t,x)
\notag\\
   &=\left[1+\frac{2(D-1)}{(D-4)(D-2)}\xi(t)\right]
   \psi^a(x)D_\mu\varphi^a(x)
\notag\\
   &\qquad{}
   +\xi(t)
   \biggl\{
   \frac{2}{(D-4)(D-2)}
   \sigma_{\mu\nu}\psi^a(x)D_\nu\varphi^a(x)
\notag\\
   &\qquad\qquad\qquad{}
   +\frac{2(D-1)}{(D-4)D}
   D_\mu\psi^a(x)\varphi^a(x)
\notag\\
   &\qquad\qquad\qquad{}
   +\frac{-2}{(D-4)D}
   \sigma_{\mu\nu}D_\nu\psi^a(x)\varphi^a(x)
   \biggr\}
\notag\\
   &\qquad{}
   +\xi(t)
   \biggl\{
   \frac{4}{(D-4)D}
   P_-\psi^a(x)D_\mu\varphi^a(x)
\notag\\
   &\qquad\qquad\qquad{}
   +\frac{8}{(D-4)(D-2)D}
   \sigma_{\mu\nu}P_-\psi^a(x)D_\nu\varphi^a(x)
\notag\\
   &\qquad\qquad\qquad{}
   +\frac{4}{(D-4)(D-2)}
   P_-D_\mu\psi^a(x)\varphi^a(x)
\notag\\
   &\qquad\qquad\qquad{}
   +\frac{-4}{(D-2)(D+2)}
   P_-\psi^a(x)D_\mu\left[\varphi^a(x)+\varphi^{\dagger a}(x)\right]
\notag\\
   &\qquad\qquad\qquad{}
   +\frac{-4(D+4)}{(D-2)D(D+2)}
   P_-D_\mu\psi^a(x)\left[\varphi^a(x)+\varphi^{\dagger a}(x)\right]
   \biggr\}
\notag\\
   &\qquad{}
   +\xi(t)\sqrt{2}
   \biggl\{
   \frac{-2}{(D-4)(D-2)D}\frac{1}{g_0}
   \gamma_\mu\sigma_{\rho\sigma}P_-\psi^a(x)F_{\rho\sigma}^a(x)
\notag\\
   &\qquad\qquad\qquad\qquad{}
   +\frac{8}{(D-4)(D-2)D}
   \frac{1}{g_0}\gamma_\nu P_-\psi^a(x)F_{\mu\nu}^a(x)
\notag\\
   &\qquad\qquad\qquad\qquad{}
   +\frac{-2(D+4)}{(D-4)(D-2)D}
   g_0f^{abc}\gamma_\mu P_-\psi^a(x)\varphi^{\dagger b}(x)\varphi^c(x)
\notag\\
   &\qquad\qquad\qquad\qquad{}
   +\frac{-2}{(D-2)D}
   g_0f^{abc}\gamma_\mu\gamma_5\psi^a(x)\varphi^{\dagger b}(x)\varphi^c(x)
   \biggr\}
   +O(t).
\label{eq:(8.22)}
\end{align}

From diagrams B01, B04, B06, B08, B10, and~B12, we have
\begin{align}
   \chi^a(t,x)\phi^a(t,x)
   &=\left[1+\frac{4(D-1)}{(D-4)(D-2)}\xi(t)\right]
   \psi^a(x)\varphi^a(x)
\notag\\
   &\qquad{}
   +\xi(t)\biggl\{
   \frac{8}{(D-4)(D-2)}
   P_-\psi^a(x)\varphi^a(x)
\notag\\
   &\qquad\qquad\qquad{}
   +\frac{-8}{(D-2)D}
   P_-\psi^a(x)\left[\varphi^a(x)+\varphi^{\dagger a}(x)\right]
   \biggr\}
   +O(t).
\label{eq:(8.23)}
\end{align}
Using the relation
$\partial_\mu(\chi^a\phi^a)=(D_\mu\chi^a)\phi^a+\chi^aD_\mu\phi^a$, we can also
deduce the small flow time expansion of $(D_\mu\chi^a)\phi^a$
from~Eqs.~\eqref{eq:(8.22)} and~\eqref{eq:(8.23)}.

On the other hand, by applying the parity transformations
of~Eqs.~\eqref{eq:(B1)}--\eqref{eq:(B7)} to this, we infer that
\begin{align}
   \chi^a(t,x)\phi^{\dagger a}(t,x)
   &=\left[1+\frac{4(D-1)}{(D-4)(D-2)}\xi(t)\right]
   \psi^a(x)\varphi^{\dagger a}(x)
\notag\\
   &\qquad{}
   +\xi(t)\biggl\{
   \frac{8}{(D-4)(D-2)}
   P_+\psi^a(x)\varphi^{\dagger a}(x)
\notag\\
   &\qquad\qquad\qquad{}
   +\frac{-8}{(D-2)D}
   P_+\psi^a(x)\left[\varphi^a(x)+\varphi^{\dagger a}(x)\right]
   \biggr\}
   +O(t).
\label{eq:(8.24)}
\end{align}

Finally, for~$g_0f^{abc}\chi^a(t,x)\phi^{\dagger b}(t,x)\phi^c(t,x)$, diagrams
D01--D11 give rise to
\begin{align}
   &g_0f^{abc}\chi^a(t,x)\phi^{\dagger b}(t,x)\phi^c(t,x)
\notag\\
   &=\left[1+\frac{2(3D^2-6D-8)}{(D-4)(D-2)D}\xi(t)\right]
   g_0f^{abc}\psi^a(x)\varphi^{\dagger b}(x)\varphi^c(x)
\notag\\
   &\qquad{}
   +\xi(t)\sqrt{2}
   \frac{2}{(D-4)(D-2)}
   \gamma_\mu
   \left[
   P_+D_\mu\psi^a(x)\varphi^a(x)+P_-D_\mu\psi^a(x)\varphi^{\dagger a}(x)
   \right]
   +O(t).
\label{eq:(8.25)}
\end{align}

It is easy to invert the above relations and obtain expressions for composite
operators of the unflowed fields in terms of composite operators of flowed
fields to the one-loop order. For example, Eq.~\eqref{eq:(8.23)}
yields
\begin{align}
   \psi^a(x)\varphi^a(x)
   &=\left[1+\frac{-4(D-1)}{(D-4)(D-2)}\xi(t)\right]
   \chi^a(t,x)\phi^a(t,x)
\notag\\
   &\qquad{}
   +\xi(t)\biggl\{
   \frac{-8}{(D-4)(D-2)}
   P_-\chi^a(t,x)\phi^a(t,x)
\notag\\
   &\qquad\qquad\qquad{}
   +\frac{8}{(D-2)D}
   P_-\chi^a(t,x)\left[\phi^a(t,x)+\phi^{\dagger a}(t,x)\right]
   \biggr\}+O(t).
\label{eq:(8.26)}
\end{align}
Similar inversions can be made for other relations.


\subsection{Final steps}
We substitute the relations in the small flow time expansion presented in the
last subsection into the expression of the supercurrent, Eq.~\eqref{eq:(2.20)}
[in this form, we do not need Eq.~\eqref{eq:(8.25)}]. Then we rewrite the
expression in terms of the renormalized gauge coupling in~Eq.~\eqref{eq:(5.3)}
and the ringed fields in~Sect.~\ref{sec:8.2}. Taking the limit~$D\to4$, we
finally find
\begin{align}
   &\Tilde{S}_\mu^{\text{imp}}
\notag\\
   &=\left\{
   1+\frac{g^2}{(4\pi)^2}C_2(G)
   \left[-\ln(8\pi\mu^2 t)-\frac{9}{4}+\frac{1}{2}\ln(432)\right]
   \right\}\left(-\frac{1}{4g}\right)
   \sigma_{\rho\sigma}\gamma_\mu\mathring{\chi}^aG_{\rho\sigma}^a
\notag\\
   &\qquad{}
   -\frac{g}{(4\pi)^2}C_2(G)
   \gamma_\nu\mathring{\chi}^a
   G_{\nu\mu}^a
\notag\\
   &\qquad{}
   +\left\{1+\frac{g^2}{(4\pi)^2}C_2(G)
   \left[-\frac{19}{4}+4\ln2+\frac{1}{2}\ln(432)\right]\right\}
\notag\\
   &\qquad\qquad\qquad{}
   \times\frac{1}{2\sqrt{2}}
   \left(\frac{1}{3}\sigma_{\mu\nu}-\delta_{\mu\nu}\right)
   (P_+D_\nu\mathring{\chi}^a\mathring{\phi}^a
   -P_-D_\nu\mathring{\chi}^a\mathring{\phi}^{\dagger a})
\notag\\
   &\qquad{}
   -\frac{3}{\sqrt{2}}\frac{g^2}{(4\pi)^2}C_2(G)
   (P_+D_\mu\mathring{\chi}^a\mathring{\phi}^a
   -P_-D_\mu\mathring{\chi}^a\mathring{\phi}^{\dagger a})
\notag\\
   &\qquad{}
   +\left\{1+\frac{g^2}{(4\pi)^2}C_2(G)
   \left[\frac{1}{2}+4\ln2+\frac{1}{2}\ln(432)\right]\right\}
\notag\\
   &\qquad\qquad\qquad{}
   \times\left(-\frac{1}{\sqrt{2}}\right)
   \left(\frac{1}{3}\sigma_{\mu\nu}-\delta_{\mu\nu}\right)
   (P_+\mathring{\chi}^aD_\nu\mathring{\phi}^a
   -P_-\mathring{\chi}^aD_\nu\mathring{\phi}^{\dagger a})
\notag\\
   &\qquad{}
   +\frac{1}{\sqrt{2}}\frac{g^2}{(4\pi)^2}C_2(G)
   \left(\frac{1}{3}\sigma_{\mu\nu}-\delta_{\mu\nu}\right)
   \gamma_5D_\nu\mathring{\chi}^a(\mathring{\phi}^a+\mathring{\phi}^{\dagger a})
\notag\\
   &\qquad{}
   +\frac{1}{2\sqrt{2}}\frac{g^2}{(4\pi)^2}C_2(G)
   \left(\frac{1}{3}\sigma_{\mu\nu}-\delta_{\mu\nu}\right)
   \gamma_5\mathring{\chi}^aD_\nu(\mathring{\phi}^a+\mathring{\phi}^{\dagger a})
\notag\\
   &\qquad{}
   -\frac{1}{4}\frac{g^3}{(4\pi)^2}C_2(G)
   f^{abc}\gamma_5\gamma_\mu
   \mathring{\chi}^a\mathring{\phi}^{\dagger b}\mathring{\phi}^c+O(t).
\label{eq:(8.27)}
\end{align}

The conjugate supercurrent $\Tilde{\Bar{S}}_\mu^{\text{imp}}$ is related
to~$\Tilde{S}_\mu^{\text{imp}}$ by the charge conjugation
in~Eqs.~\eqref{eq:(B8)}--\eqref{eq:(B16)} as~$\Tilde{S}_\mu^{\text{imp}}\to%
C(\Tilde{\Bar{S}}_\mu^{\text{imp}})^T$. Using this, we have
\begin{align}
   &\Tilde{\Bar{S}}_\mu^{\text{imp}}
\notag\\
   &=\left\{
   1+\frac{g^2}{(4\pi)^2}C_2(G)
   \left[-\ln(8\pi\mu^2 t)-\frac{9}{4}+\frac{1}{2}\ln(432)\right]
   \right\}\left(-\frac{1}{4g}\right)
   \mathring{\Bar{\chi}}^a\gamma_\mu\sigma_{\rho\sigma}G_{\rho\sigma}^a
\notag\\
   &\qquad{}
   +\frac{g}{(4\pi)^2}C_2(G)
   \mathring{\Bar{\chi}}^a\gamma_\nu
   G_{\nu\mu}^a
\notag\\
   &\qquad{}
   +\left\{1+\frac{g^2}{(4\pi)^2}C_2(G)
   \left[-\frac{19}{4}+4\ln2+\frac{1}{2}\ln(432)\right]\right\}
\notag\\
   &\qquad\qquad\qquad{}
   \times\left(-\frac{1}{2\sqrt{2}}\right)
   (D_\nu\mathring{\Bar{\chi}}^aP_+\mathring{\phi}^a
   -D_\nu\mathring{\Bar{\chi}}^aP_-\mathring{\phi}^{\dagger a})
   \left(\frac{1}{3}\sigma_{\nu\mu}-\delta_{\nu\mu}\right)
\notag\\
   &\qquad{}
   +\frac{3}{\sqrt{2}}\frac{g^2}{(4\pi)^2}C_2(G)
   (D_\mu\mathring{\Bar{\chi}}^aP_+\mathring{\phi}^a
   -D_\mu\mathring{\Bar{\chi}}^aP_-\mathring{\phi}^{\dagger a})
\notag\\
   &\qquad{}
   +\left\{1+\frac{g^2}{(4\pi)^2}C_2(G)
   \left[\frac{1}{2}+4\ln2+\frac{1}{2}\ln(432)\right]\right\}
\notag\\
   &\qquad\qquad\qquad{}
   \times\frac{1}{\sqrt{2}}
   (P_+\mathring{\Bar{\chi}}^aD_\nu\mathring{\phi}^a
   -P_-\mathring{\Bar{\chi}}^aD_\nu\mathring{\phi}^{\dagger a})
   \left(\frac{1}{3}\sigma_{\nu\mu}-\delta_{\nu\mu}\right)
\notag\\
   &\qquad{}
   -\frac{1}{\sqrt{2}}\frac{g^2}{(4\pi)^2}C_2(G)
   D_\nu\mathring{\Bar{\chi}}^a\gamma_5
   (\mathring{\phi}^a+\mathring{\phi}^{\dagger a})
   \left(\frac{1}{3}\sigma_{\nu\mu}-\delta_{\nu\mu}\right)
\notag\\
   &\qquad{}
   -\frac{1}{2\sqrt{2}}\frac{g^2}{(4\pi)^2}C_2(G)
   \mathring{\Bar{\chi}}^a\gamma_5
   D_\nu(\mathring{\phi}^a+\mathring{\phi}^{\dagger a})
   \left(\frac{1}{3}\sigma_{\nu\mu}-\delta_{\nu\mu}\right)
\notag\\
   &\qquad{}
   +\frac{1}{4}\frac{g^3}{(4\pi)^2}C_2(G)
   f^{abc}
   \mathring{\Bar{\chi}}^a\gamma_\mu\gamma_5
   \mathring{\phi}^{\dagger b}\mathring{\phi}^c+O(t).
\label{eq:(8.28)}
\end{align}

Some remarks are in order: (1)~The composite operators
in~Eqs.~\eqref{eq:(8.27)} and~\eqref{eq:(8.28)} are written completely in terms
of the renormalized gauge coupling, the flowed gauge field, and the ringed
flowed fermion and scalar fields. Thus, these operators are \emph{manifestly
finite\/} renormalized operators being independent of the regularization.
(2)~From these expressions, we have the $\gamma$-trace anomaly:\footnote{Note
that we can set $\gamma_\mu\sigma_{\rho\sigma}\gamma_\mu=%
\gamma_\mu[(1/3)\sigma_{\mu\nu}-\delta_{\mu\nu}]=0$ in these finite (and thus
$D=4$) expressions}
\begin{align}
   \gamma_\mu\Tilde{S}_\mu^{\text{imp}}
   &=\frac{g}{(4\pi)^2}C_2(G)\sigma_{\mu\nu}\mathring{\chi}^aG_{\mu\nu}^a
   -\frac{3}{\sqrt{2}}\frac{g^2}{(4\pi)^2}C_2(G)
   (P_-\Slash{D}\mathring{\chi}^a\mathring{\phi}^a
   -P_+\Slash{D}\mathring{\chi}^a\mathring{\phi}^{\dagger a})
\notag\\
   &\qquad{}
   +\frac{g^3}{(4\pi)^2}C_2(G)f^{abc}
   \gamma_5\mathring{\chi}^a\mathring{\phi}^{\dagger b}\mathring{\phi}^c+O(t)
\notag\\
   &=\frac{g}{(4\pi)^2}C_2(G)\sigma_{\mu\nu}\mathring{\chi}^aG_{\mu\nu}^a
   -2\frac{g^3}{(4\pi)^2}C_2(G)f^{abc}
   \gamma_5\mathring{\chi}^a\mathring{\phi}^{\dagger b}\mathring{\phi}^c+O(t),
\label{eq:(8.29)}
\end{align}
where in the last line we have used equations of motion whose use is justified
in on-mass-shell correlation functions. Note that in the tree level, the flowed
fields are identical to unflowed ones up to~$O(t)$ terms. (3)~To the one-loop
order, Eqs.~\eqref{eq:(8.27)} and~\eqref{eq:(8.28)} give the properly
normalized supercurrent, as argued in~Eq.~\eqref{eq:(7.6)}. The difference
between this approximation and the ``would-be true supercurrent'' will
be~$O(g_0^2)$. Since we will invoke the renormalization group improvement to be
shortly discussed, this difference can be neglected in the final expressions in
the~$t\to0$ limit. (4)~Since the composite operators on both sides
of~Eq.~\eqref{eq:(8.27)},
$\Tilde{S}_\mu^{\text{imp}}(x)$~[Eq.~\eqref{eq:(2.20)}],
$\mathcal{O}_1(t,x)\equiv%
\sigma_{\rho\sigma}\gamma_\mu\mathring{\chi}^a(t,x)G_{\rho\sigma}^a(t,x)$, etc.\
are bare operators, the derivative of the coefficients~$c_1(t)$ etc., where
$\Tilde{S}_\mu^{\text{imp}}(x)\equiv c_1(t)\mathcal{O}_1(t,x)+\dotsb$, with
respect to the renormalization scale~$\mu$ while bare quantities are kept fixed
vanishes:
\begin{equation}
   \left(\mu\frac{\partial}{\partial\mu}\right)_0c_1(t)=0.
\label{eq:(8.30)}
\end{equation}
By the standard argument, this implies that we can set the renormalization
scale~$\mu$ in~$c_1(t)$ arbitrarily, if the renormalized gauge coupling~$g$
in~$c_1(t)$ is replaced by the running gauge coupling~$\Bar{g}(\mu)$ defined by
\begin{equation}
   \mu\frac{d\Bar{g}(\mu)}{d\mu}=\beta(\Bar{g}(\mu)),
\end{equation}
where the beta function is~$\beta(g)=-2g^3C_2(G)/(4\pi)^2$ to all orders in
perturbation theory~\cite{Tarasov:1980au,Avdeev:1981ew,Grisaru:1982zh,%
Gates:1983nr,Seiberg:1988ur}. In fact,
\begin{align}
   \frac{d}{d\mu}\left.c_1(t)\right|_{g=\Bar{g}(\mu)}
   &=\left[\mu\frac{\partial}{\partial\mu}
   +\mu\frac{d\Bar{g}(\mu)}{d\mu}\frac{\partial}{\partial\Bar{g}(\mu)}\right]
   \left.c_1(t)\right|_{g=\Bar{g}(\mu)}
\notag\\
   &=\left[\mu\frac{\partial}{\partial\mu}
   +\beta(\Bar{g}(\mu))\frac{\partial}{\partial\Bar{g}(\mu)}\right]
   \left.c_1(t)\right|_{g=\Bar{g}(\mu)}
\notag\\
   &=\left.\left[\mu\frac{\partial}{\partial\mu}
   +\beta(g)\frac{\partial}{\partial g}\right]c_1(t)\right|_{g=\Bar{g}(\mu)}
\notag\\
   &=\left.\left(\mu\frac{\partial}{\partial\mu}\right)_0c_1(t)
   \right|_{g=\Bar{g}(\mu)}
\notag\\
   &=0.
\end{align}
Here, we have used the definition of the beta function
\begin{equation}
   \beta(g)=\left(\mu\frac{\partial}{\partial\mu}\right)_0g
\end{equation}
and Eq.~\eqref{eq:(8.30)}. We thus take $\mu=1/\sqrt{8t}$ by using the flow
time~$t$. Then, taking the $t\to0$ limit, we have our formulas,
Eqs.~\eqref{eq:(1.1)} and~\eqref{eq:(1.2)}. This completes our argument.

\section*{Acknowledgments}
We would like to thank Hiroki Makino for discussions.
This work was supported by JSPS Grant-in-Aid for Scientific Research Grant
Numbers, JP16J02259 (A.~K.), JP18J20935 (O.~M.), and JP16H03982 (H.~S.).

\appendix

\section{Notational convention}
\label{sec:A}
Without noting otherwise, repeated indices are understood to be summed over.
The spacetime dimension is denoted by~$D\equiv4-2\epsilon$. We use the
following abbreviation for the momentum integral:
\begin{equation}
   \int_p\equiv\int\frac{d^Dp}{(2\pi)^D}.
\label{eq:(A1)}
\end{equation}

Our Dirac matrices~$\gamma_\mu$,
satisfying~$\{\gamma_\mu,\gamma_\nu\}=2\delta_{\mu\nu}$, are all Hermitian and
for the trace over the spinor index we set $\tr(1)=4$ for any spacetime
dimension~$D$. The chiral matrix and chirality projection operators are
defined by
\begin{align}
   \gamma_5\equiv\gamma_0\gamma_1\gamma_2\gamma_3,\qquad
   P_\pm\equiv\frac{1}{2}(1\pm\gamma_5),
\label{eq:(A2)}
\end{align}
for \emph{any}~$D$; we have
\begin{equation}
   \tr(\gamma_5\gamma_\mu\gamma_\nu\gamma_\rho\gamma_\sigma)
   =\begin{cases}
   4\epsilon_{\mu\nu\rho\sigma},&\mu,\nu,\rho,\sigma\in\{0,1,2,3\},\\
   0,&\text{otherwise},
   \end{cases}
\label{eq:(A3)}
\end{equation}
where the totally anti-symmetric tensor is normalized as~$\epsilon_{0123}=1$.
We also use
\begin{equation}
   \sigma_{\mu\nu}\equiv\frac{1}{2}[\gamma_\mu,\gamma_\nu].
\label{eq:(A4)}
\end{equation}

In the 4D $\mathcal{N}=2$ SYM, all fields belong to the adjoint representation
and the covariant derivatives for a generic field~$X$ are defined from the
structure constants of the gauge group~$f^{abc}$ by 
\begin{align}
   D_\mu X^a
   &\equiv\partial_\mu X^a+f^{abc}A_\mu^b X^c,
\label{eq:(A5)}
\\
   X^a\overleftarrow{D}_\mu
   &\equiv X^a\overleftarrow{\partial}_\mu-X^cf^{cba}A_\mu^b.
\label{eq:(A6)}
\end{align}
We also use the abbreviations $\Slash{D}\equiv\gamma_\mu D_\mu$
and~$\overleftarrow{\Slash{D}}\equiv\gamma_\mu\overleftarrow{D}_\mu$. The
quadratic Casimir~$C_2(G)$ is defined by~$f^{bXa}f^{aYb}=-C_2(G)\delta^{XY}$. A
useful identity is
\begin{equation}
   f^{cXa}f^{aYb}f^{bZc}=-\frac{1}{2}C_2(G)f^{XYZ}.
\label{eq:(A7)}
\end{equation}

\section{The parity and charge conjugation invariance}
\label{sec:B}
The actions, the flow equations and the initial conditions, all elements of the
present system are invariant under the following parity transformation and
charge conjugation.

The parity transformation is defined, denoting the spatial directions $\mu=1$,
$2$, and~$3$ by~$i$, by
\begin{align}
   \psi(x)&\to\gamma_0\psi(\Tilde x),&
   \Bar{\psi}(x)&\to\Bar{\psi}(\Tilde x)\gamma_0,
\label{eq:(B1)}
\\
   A_0(x)&\to A_0(\Tilde x),&
   A_i(x)&\to-A_i(\Tilde x),
\label{eq:(B2)}
\\
   \varphi(x)&\to-\varphi^\dagger(\Tilde x),&
   \varphi^\dagger(x)&\to-\varphi(\Tilde x),
\label{eq:(B3)}
\\
   c(x)&\to c(\Tilde x),&
   \Bar{c}(x)&\to\Bar{c}(\Tilde x),
\label{eq:(B4)}
\end{align}
where $\Tilde x\equiv(x_0,-x_i)$ and
\begin{align}
   \chi(t,x)&\to\gamma_0\chi(t,\Tilde x),&
   \Bar{\chi}(t,x)&\to\Bar{\chi}(t,\Tilde x)\gamma_0,
\label{eq:(B5)}
\\
   B_0(t,x)&\to A_0(t,\Tilde x),&
   B_i(t,x)&\to-A_i(t,\Tilde x),
\label{eq:(B6)}
\\
   \phi(t,x)&\to-\phi^\dagger(t,\Tilde x),&
   \phi^\dagger(t,x)&\to-\phi(t,\Tilde x).
\label{eq:(B7)}
\end{align}

The charge conjugation, on the other hand, is defined by
\begin{align}
   \psi(x)&\to C\Bar{\psi}^T(x),&
   \Bar{\psi}(x)&\to-\psi^T(x)C^{-1},
\label{eq:(B8)}
\\
   A_\mu(x)&\to A_\mu(x),&
\label{eq:(B9)}
\\
   \varphi(x)&\to-\varphi(x),&
   \varphi^\dagger(x)&\to-\varphi^\dagger(x),
\label{eq:(B10)}
\\
   c(x)&\to c(x),&
   \Bar{c}(x)&\to\Bar{c}(x),
\label{eq:(B11)}
\end{align}
where $C$ is the charge conjugation matrix satisfying
\begin{equation}
   C^{-1}\gamma_\mu C=-\gamma_\mu^T,\qquad C^T=-C,
\label{eq:(B12)}
\end{equation}
and thus
\begin{equation}
   C^{-1}\sigma_{\mu\nu}C=-\sigma_{\mu\nu}^T,\qquad
   C^{-1}\gamma_5C=\gamma_5^T,
\label{eq:(B13)}
\end{equation}
for any~$D$; see Appendix~A of~Ref.~\cite{Hieda:2017sqq}. Correspondingly,
\begin{align}
   \chi(t,x)&\to C\Bar{\chi}^T(t,x),&
   \Bar{\chi}(t,x)&\to-\chi^T(t,x)C^{-1},
\label{eq:(B14)}
\\
   B_\mu(t,x)&\to B_\mu(t,x),&
\label{eq:(B15)}
\\
   \phi(t,x)&\to-\phi(t,x),&
   \phi^\dagger(t,x)&\to-\phi^\dagger(t,x).
\label{eq:(B16)}
\end{align}

\section{(Flow) Feynman diagrams}
\label{sec:C}
Here we collect Feynman diagrams and flow Feynman
diagrams~\cite{Luscher:2010iy,Luscher:2011bx} that are relevant to the
computations in the main text. We basically follow the convention
in~Ref.~\cite{Makino:2014taa}: The wavy line and the straight arrowed line
represent the gauge field propagator and the Dirac fermion field propagator,
respectively. In addition to these, in this paper the broken line represents
the scalar field propagator. Doubled lines denote the corresponding heat
kernels~\cite{Luscher:2011bx,Luscher:2013cpa,Makino:2014taa}. That is, the
double wavy line, the double arrowed line, and the double broken line represent
the gauge field heat kernel, the fermion field heat kernel, and the scalar
field heat kernel, respectively. The black bullet denotes the interaction
vertex in the original action, while the white circle denotes the interaction
term in the flow
equations~\cite{Luscher:2011bx,Luscher:2013cpa,Makino:2014taa}. The x-mark
generally represents the composite operator under consideration.

\begin{figure}[htbp]
\centering
\begin{subfigure}{0.24\columnwidth}
\centering
\includegraphics[width=0.7\columnwidth]{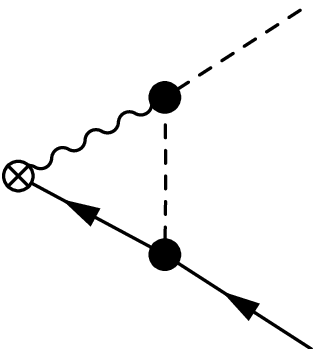}
\caption{A01}
\end{subfigure}
\begin{subfigure}{0.24\columnwidth}
\centering
\includegraphics[width=0.7\columnwidth]{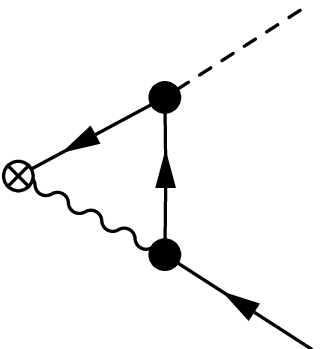}
\caption{A02}
\end{subfigure}
\caption{One-loop (flow) Feynman diagrams for the operator renormalization and
the small flow time expansion.}
\label{fig:C1}
\end{figure}

\begin{figure}[htbp]
\centering
\begin{subfigure}{0.24\columnwidth}
\centering
\includegraphics[width=0.7\columnwidth]{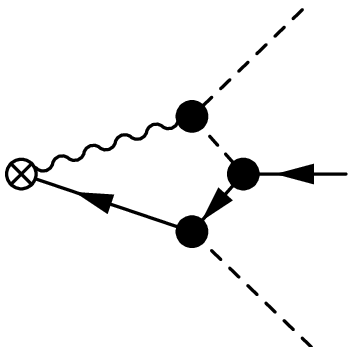}
\caption{A03}
\end{subfigure}
\begin{subfigure}{0.24\columnwidth}
\centering
\includegraphics[width=0.7\columnwidth]{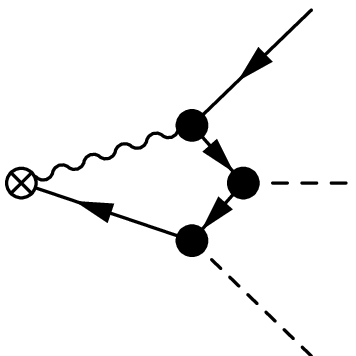}
\caption{A04}
\end{subfigure}
\begin{subfigure}{0.24\columnwidth}
\centering
\includegraphics[width=0.7\columnwidth]{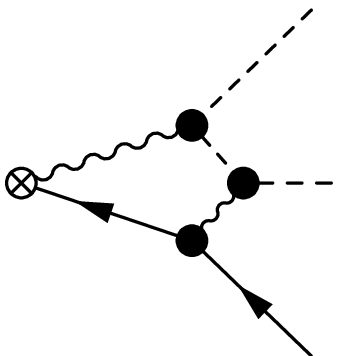}
\caption{A05}
\end{subfigure}
\begin{subfigure}{0.24\columnwidth}
\centering
\includegraphics[width=0.7\columnwidth]{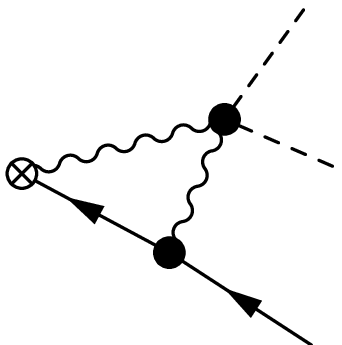}
\caption{A06}
\end{subfigure}
\caption{One-loop (flow) Feynman diagrams (continued).}
\label{fig:C2}
\end{figure}

\begin{figure}[htbp]
\centering
\begin{subfigure}{0.24\columnwidth}
\centering
\includegraphics[width=0.7\columnwidth]{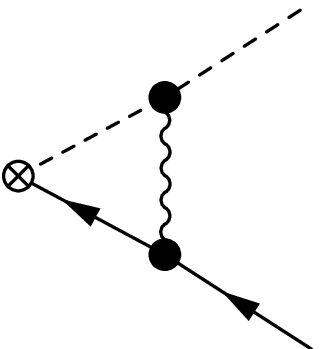}
\caption{B01}
\end{subfigure}
\begin{subfigure}{0.24\columnwidth}
\centering
\includegraphics[width=0.7\columnwidth]{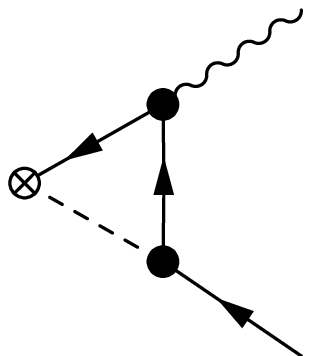}
\caption{B02}
\end{subfigure}
\begin{subfigure}{0.24\columnwidth}
\centering
\includegraphics[width=0.7\columnwidth]{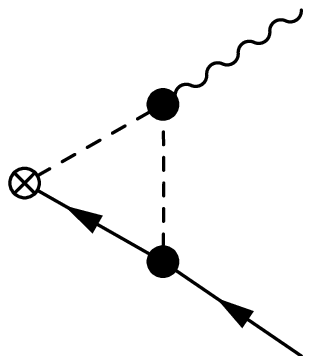}
\caption{B03}
\end{subfigure}
\begin{subfigure}{0.24\columnwidth}
\centering
\includegraphics[width=0.7\columnwidth]{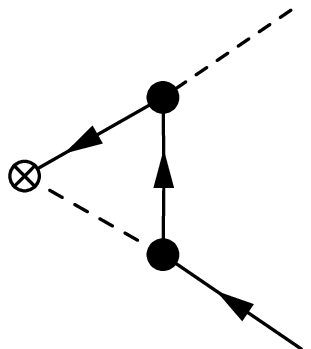}
\caption{B04}
\end{subfigure}
\begin{subfigure}{0.24\columnwidth}
\centering
\includegraphics[width=0.7\columnwidth]{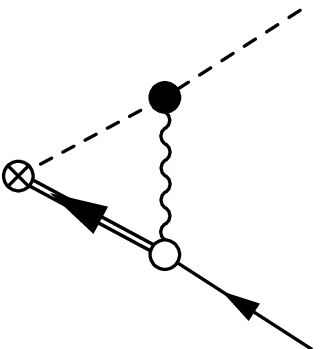}
\caption{B05}
\end{subfigure}
\begin{subfigure}{0.24\columnwidth}
\centering
\includegraphics[width=0.7\columnwidth]{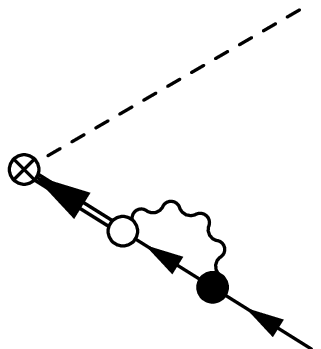}
\caption{B06}
\end{subfigure}
\begin{subfigure}{0.24\columnwidth}
\centering
\includegraphics[width=0.7\columnwidth]{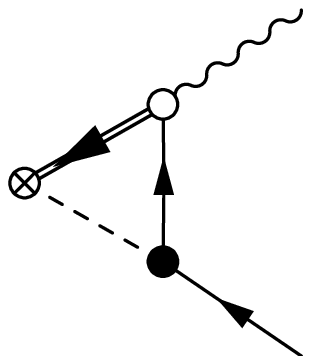}
\caption{B07}
\end{subfigure}
\begin{subfigure}{0.24\columnwidth}
\centering
\includegraphics[width=0.7\columnwidth]{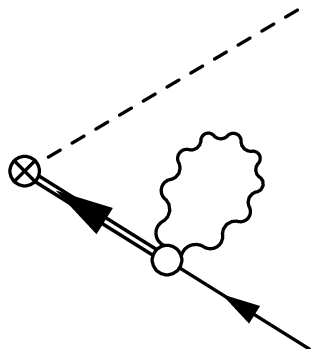}
\caption{B08}
\end{subfigure}
\begin{subfigure}{0.24\columnwidth}
\centering
\includegraphics[width=0.7\columnwidth]{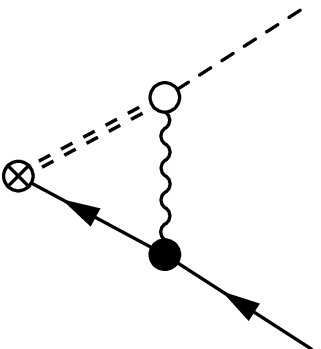}
\caption{B09}
\end{subfigure}
\begin{subfigure}{0.24\columnwidth}
\centering
\includegraphics[width=0.7\columnwidth]{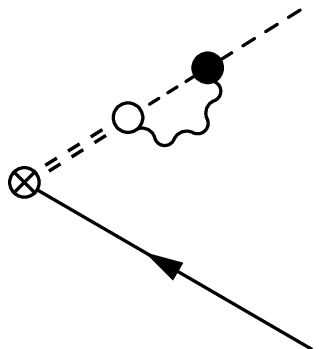}
\caption{B10}
\end{subfigure}
\begin{subfigure}{0.24\columnwidth}
\centering
\includegraphics[width=0.7\columnwidth]{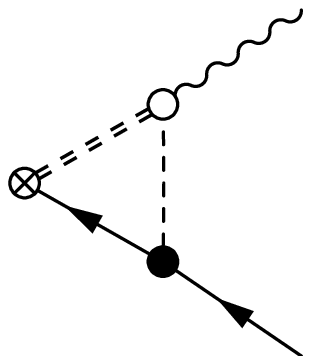}
\caption{B11}
\end{subfigure}
\begin{subfigure}{0.24\columnwidth}
\centering
\includegraphics[width=0.7\columnwidth]{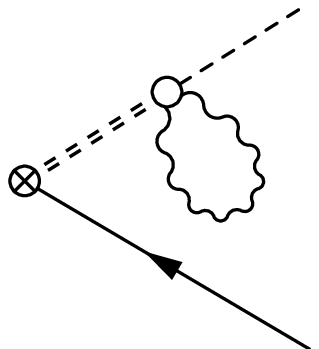}
\caption{B12}
\end{subfigure}
\begin{subfigure}{0.24\columnwidth}
\centering
\includegraphics[width=0.7\columnwidth]{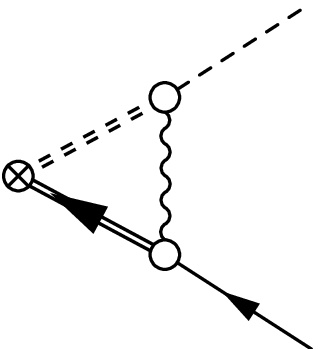}
\caption{B13}
\end{subfigure}
\begin{subfigure}{0.24\columnwidth}
\centering
\includegraphics[width=0.7\columnwidth]{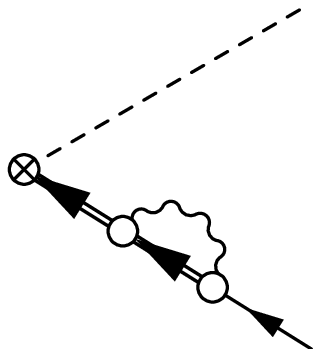}
\caption{B14}
\end{subfigure}
\begin{subfigure}{0.24\columnwidth}
\centering
\includegraphics[width=0.7\columnwidth]{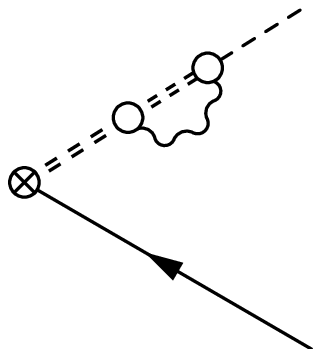}
\caption{B15}
\end{subfigure}
\caption{One-loop (flow) Feynman diagrams (continued).}
\label{fig:C3}
\end{figure}

\begin{figure}[htbp]
\centering
\begin{subfigure}{0.24\columnwidth}
\centering
\includegraphics[width=0.7\columnwidth]{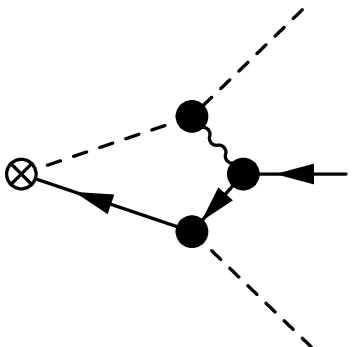}
\caption{B16}
\end{subfigure}
\begin{subfigure}{0.24\columnwidth}
\centering
\includegraphics[width=0.7\columnwidth]{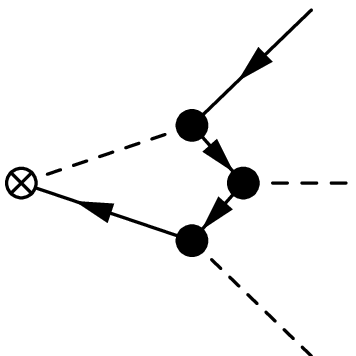}
\caption{B17}
\end{subfigure}
\begin{subfigure}{0.24\columnwidth}
\centering
\includegraphics[width=0.7\columnwidth]{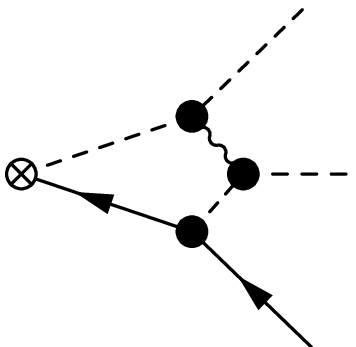}
\caption{B18}
\end{subfigure}
\begin{subfigure}{0.24\columnwidth}
\centering
\includegraphics[width=0.7\columnwidth]{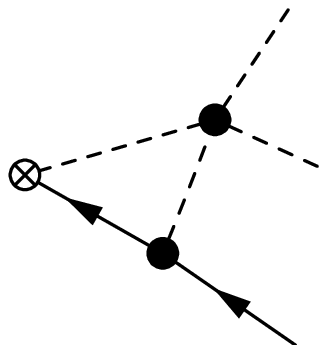}
\caption{B19}
\end{subfigure}
\begin{subfigure}{0.24\columnwidth}
\centering
\includegraphics[width=0.7\columnwidth]{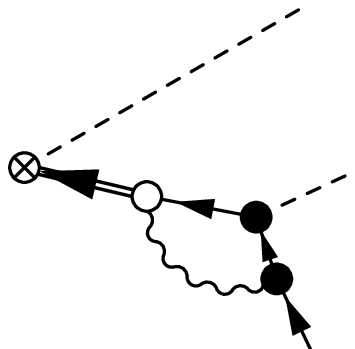}
\caption{B20}
\end{subfigure}
\caption{One-loop (flow) Feynman diagrams (continued).}
\label{fig:C4}
\end{figure}

\begin{figure}[htbp]
\centering
\begin{subfigure}{0.24\columnwidth}
\centering
\includegraphics[width=0.7\columnwidth]{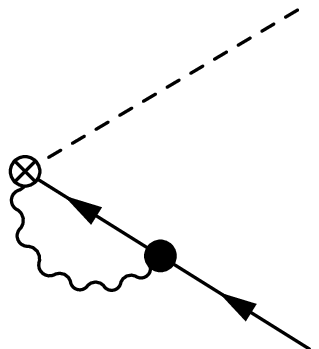}
\caption{C01}
\end{subfigure}
\begin{subfigure}{0.24\columnwidth}
\centering
\includegraphics[width=0.7\columnwidth]{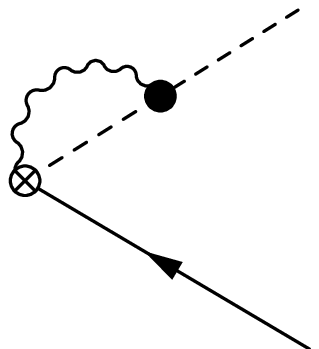}
\caption{C02}
\end{subfigure}
\begin{subfigure}{0.24\columnwidth}
\centering
\includegraphics[width=0.7\columnwidth]{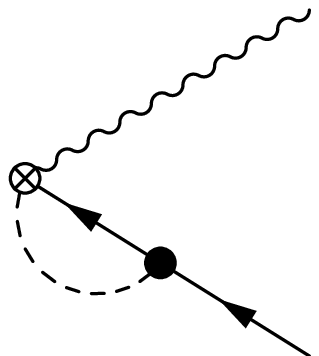}
\caption{C03}
\end{subfigure}
\begin{subfigure}{0.24\columnwidth}
\centering
\includegraphics[width=0.7\columnwidth]{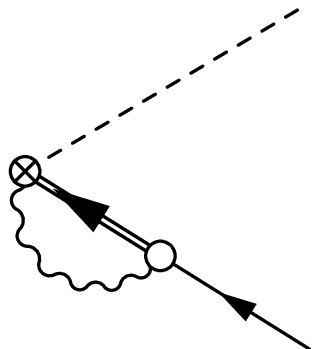}
\caption{C04}
\end{subfigure}
\begin{subfigure}{0.24\columnwidth}
\centering
\includegraphics[width=0.7\columnwidth]{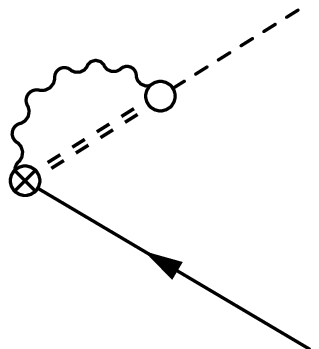}
\caption{C05}
\end{subfigure}
\caption{One-loop (flow) Feynman diagrams (continued).}
\label{fig:C5}
\end{figure}

\begin{figure}[htbp]
\centering
\begin{subfigure}{0.24\columnwidth}
\centering
\includegraphics[width=0.7\columnwidth]{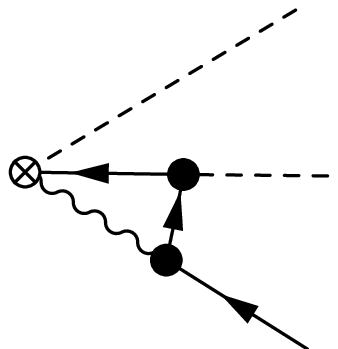}
\caption{C06}
\end{subfigure}
\begin{subfigure}{0.24\columnwidth}
\centering
\includegraphics[width=0.7\columnwidth]{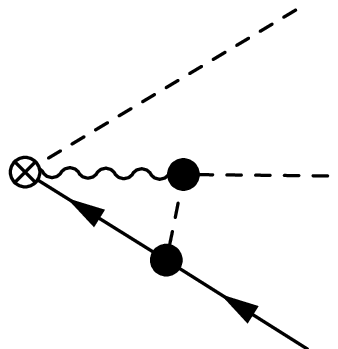}
\caption{C07}
\end{subfigure}
\caption{One-loop (flow) Feynman diagrams (continued).}
\label{fig:C6}
\end{figure}

\begin{figure}[htbp]
\centering
\begin{subfigure}{0.24\columnwidth}
\centering
\includegraphics[width=0.7\columnwidth]{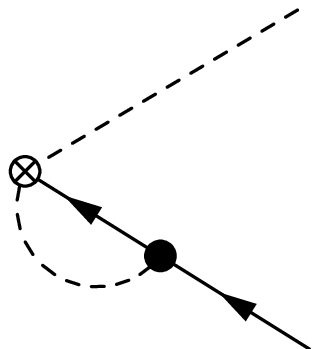}
\caption{D01}
\end{subfigure}
\begin{subfigure}{0.24\columnwidth}
\centering
\includegraphics[width=0.7\columnwidth]{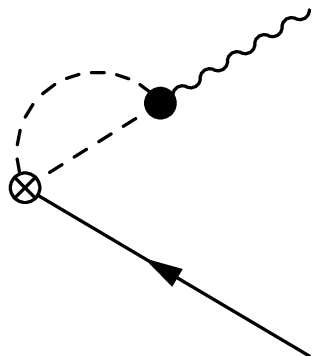}
\caption{D02}
\end{subfigure}
\begin{subfigure}{0.24\columnwidth}
\centering
\includegraphics[width=0.7\columnwidth]{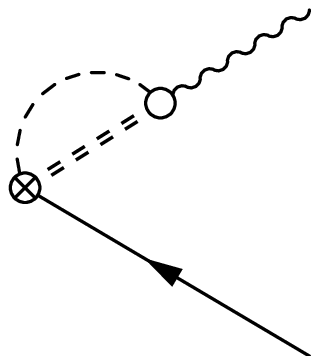}
\caption{D03}
\end{subfigure}
\caption{One-loop (flow) Feynman diagrams (continued).}
\label{fig:C7}
\end{figure}

\begin{figure}[htbp]
\centering
\begin{subfigure}{0.24\columnwidth}
\centering
\includegraphics[width=0.7\columnwidth]{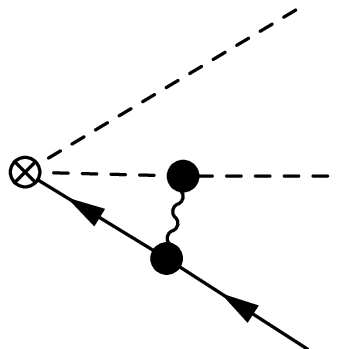}
\caption{D04}
\end{subfigure}
\begin{subfigure}{0.24\columnwidth}
\centering
\includegraphics[width=0.7\columnwidth]{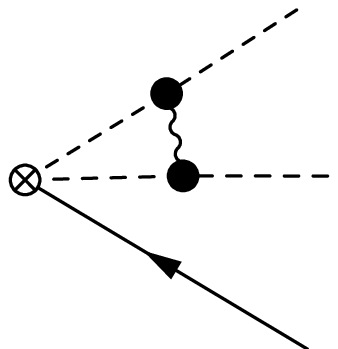}
\caption{D05}
\end{subfigure}
\begin{subfigure}{0.24\columnwidth}
\centering
\includegraphics[width=0.7\columnwidth]{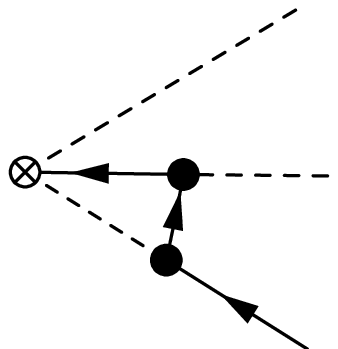}
\caption{D06}
\end{subfigure}
\begin{subfigure}{0.24\columnwidth}
\centering
\includegraphics[width=0.7\columnwidth]{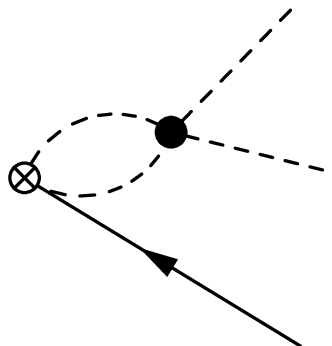}
\caption{D07}
\end{subfigure}
\begin{subfigure}{0.24\columnwidth}
\centering
\includegraphics[width=0.7\columnwidth]{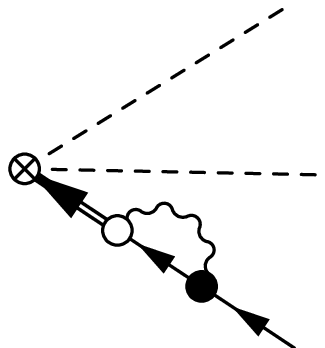}
\caption{D08}
\end{subfigure}
\begin{subfigure}{0.24\columnwidth}
\centering
\includegraphics[width=0.7\columnwidth]{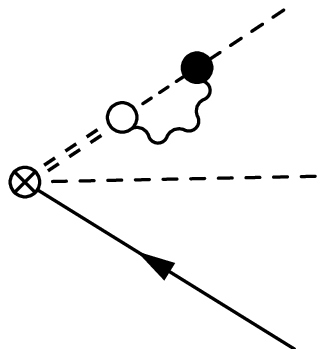}
\caption{D09}
\end{subfigure}
\begin{subfigure}{0.24\columnwidth}
\centering
\includegraphics[width=0.7\columnwidth]{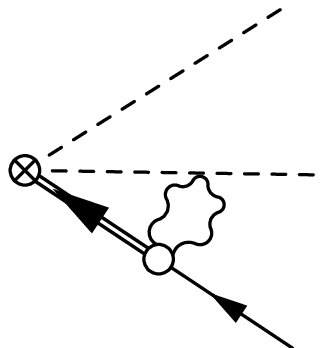}
\caption{D10}
\end{subfigure}
\begin{subfigure}{0.24\columnwidth}
\centering
\includegraphics[width=0.7\columnwidth]{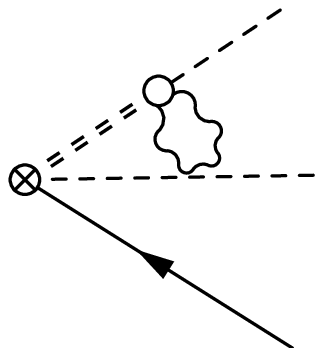}
\caption{D11}
\end{subfigure}
\caption{One-loop (flow) Feynman diagrams (continued).}
\label{fig:C8}
\end{figure}

\begin{figure}[htbp]
\centering
\begin{subfigure}{0.24\columnwidth}
\centering
\includegraphics[width=0.7\columnwidth]{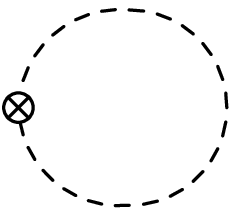}
\caption{E01}
\end{subfigure}
\begin{subfigure}{0.24\columnwidth}
\centering
\includegraphics[width=0.7\columnwidth]{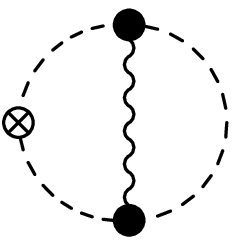}
\caption{E02}
\end{subfigure}
\begin{subfigure}{0.24\columnwidth}
\centering
\includegraphics[width=0.7\columnwidth]{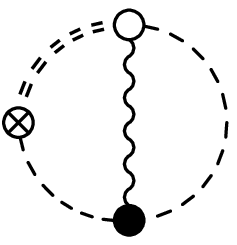}
\caption{E03}
\end{subfigure}
\begin{subfigure}{0.24\columnwidth}
\centering
\includegraphics[width=0.7\columnwidth]{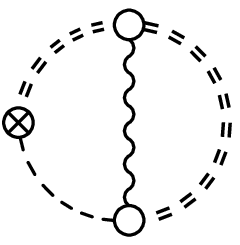}
\caption{E04}
\end{subfigure}
\begin{subfigure}{0.24\columnwidth}
\centering
\includegraphics[width=0.7\columnwidth]{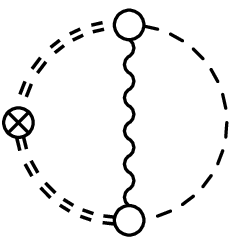}
\caption{E05}
\end{subfigure}
\begin{subfigure}{0.24\columnwidth}
\centering
\includegraphics[width=0.7\columnwidth]{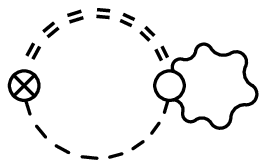}
\caption{E06}
\end{subfigure}
\begin{subfigure}{0.24\columnwidth}
\centering
\includegraphics[width=0.7\columnwidth]{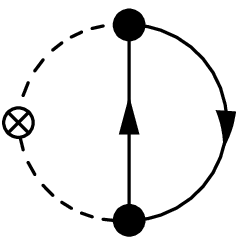}
\caption{E07}
\end{subfigure}
\caption{Flow Feynman diagrams relevant for the calculation
in~Sect.~\ref{sec:8.2}}
\label{fig:C9}
\end{figure}


\clearpage

\begin{thebibliography}{00}

\bibitem{Catterall:2009it} 
  S.~Catterall, D.~B.~Kaplan and M.~\"Unsal,
  Phys.\ Rept.\  {\bf 484}, 71 (2009)
  doi:10.1016/j.physrep.2009.09.001
  [arXiv:0903.4881 [hep-lat]].

\bibitem{Kadoh:2016eju} 
  D.~Kadoh,
  PoS LATTICE {\bf 2015}, 017 (2016)
  doi:10.22323/1.251.0017
  [arXiv:1607.01170 [hep-lat]].

\bibitem{Curci:1986sm} 
  G.~Curci and G.~Veneziano,
  Nucl.\ Phys.\ B {\bf 292}, 555 (1987).
  doi:10.1016/0550-3213(87)90660-2

\bibitem{Kaplan:1983sk} 
  D.~B.~Kaplan,
  Phys.\ Lett.\  {\bf 136B}, 162 (1984).
  doi:10.1016/0370-2693(84)91172-9

\bibitem{Farchioni:2001wx} 
  F.~Farchioni {\it et al.} [DESY-Munster-Roma Collaboration],
  Eur.\ Phys.\ J.\ C {\bf 23}, 719 (2002)
  doi:10.1007/s100520200898
  [hep-lat/0111008].

\bibitem{Suzuki:2012pc} 
  H.~Suzuki,
  Nucl.\ Phys.\ B {\bf 861}, 290 (2012)
  doi:10.1016/j.nuclphysb.2012.04.008
  [arXiv:1202.2598 [hep-lat]].

\bibitem{Ali:2018fbq} 
  S.~Ali, H.~Gerber, I.~Montvay, G.~M\"unster, S.~Piemonte, P.~Scior and G.~Bergner,
  Eur.\ Phys.\ J.\ C {\bf 78}, no. 5, 404 (2018)
  doi:10.1140/epjc/s10052-018-5887-9
  [arXiv:1802.07067 [hep-lat]].

\bibitem{Hieda:2017sqq} 
  K.~Hieda, A.~Kasai, H.~Makino and H.~Suzuki,
  PTEP {\bf 2017}, no. 6, 063B03 (2017)
  doi:10.1093/ptep/ptx073
  [arXiv:1703.04802 [hep-lat]].

\bibitem{Narayanan:2006rf} 
  R.~Narayanan and H.~Neuberger,
  JHEP {\bf 0603}, 064 (2006)
  doi:10.1088/1126-6708/2006/03/064
  [hep-th/0601210].

\bibitem{Luscher:2009eq} 
  M.~L\"uscher,
  Commun.\ Math.\ Phys.\  {\bf 293}, 899 (2010)
  doi:10.1007/s00220-009-0953-7
  [arXiv:0907.5491 [hep-lat]].

\bibitem{Luscher:2010iy} 
  M.~L\"uscher,
  JHEP {\bf 1008}, 071 (2010)
  Erratum: [JHEP {\bf 1403}, 092 (2014)]
  doi:10.1007/JHEP08(2010)071, 10.1007/JHEP03(2014)092
  [arXiv:1006.4518 [hep-lat]].

\bibitem{Luscher:2011bx} 
  M.~L\"uscher and P.~Weisz,
  JHEP {\bf 1102}, 051 (2011)
  doi:10.1007/JHEP02(2011)051
  [arXiv:1101.0963 [hep-th]].

\bibitem{Luscher:2013cpa} 
  M.~L\"uscher,
  JHEP {\bf 1304}, 123 (2013)
  doi:10.1007/JHEP04(2013)123
  [arXiv:1302.5246 [hep-lat]].

\bibitem{Luscher:2013vga} 
  M.~L\"uscher,
  PoS LATTICE {\bf 2013}, 016 (2014)
  doi:10.22323/1.187.0016
  [arXiv:1308.5598 [hep-lat]].

\bibitem{Ramos:2015dla} 
  A.~Ramos,
  PoS LATTICE {\bf 2014}, 017 (2015)
  doi:10.22323/1.214.0017
  [arXiv:1506.00118 [hep-lat]].

\bibitem{Hieda:2016xpq} 
  K.~Hieda, H.~Makino and H.~Suzuki,
  Nucl.\ Phys.\ B {\bf 918}, 23 (2017)
  doi:10.1016/j.nuclphysb.2017.02.017
  [arXiv:1604.06200 [hep-lat]].

\bibitem{Suzuki:2013gza} 
  H.~Suzuki,
  PTEP {\bf 2013}, 083B03 (2013)
  Erratum: [PTEP {\bf 2015}, 079201 (2015)]
  doi:10.1093/ptep/ptt059, 10.1093/ptep/ptv094
  [arXiv:1304.0533 [hep-lat]].

\bibitem{Makino:2014taa} 
  H.~Makino and H.~Suzuki,
  PTEP {\bf 2014}, 063B02 (2014)
  Erratum: [PTEP {\bf 2015}, 079202 (2015)]
  doi:10.1093/ptep/ptu070, 10.1093/ptep/ptv095
  [arXiv:1403.4772 [hep-lat]].

\bibitem{DelDebbio:2013zaa} 
  L.~Del Debbio, A.~Patella and A.~Rago,
  JHEP {\bf 1311}, 212 (2013)
  doi:10.1007/JHEP11(2013)212
  [arXiv:1306.1173 [hep-th]].

\bibitem{Caracciolo:1989pt} 
  S.~Caracciolo, G.~Curci, P.~Menotti and A.~Pelissetto,
  Annals Phys.\  {\bf 197}, 119 (1990).
  doi:10.1016/0003-4916(90)90203-Z

\bibitem{Suzuki:2016ytc} 
  H.~Suzuki,
  PoS LATTICE {\bf 2016}, 002 (2017)
  doi:10.22323/1.256.0002
  [arXiv:1612.00210 [hep-lat]].

\bibitem{Asakawa:2013laa} 
  M.~Asakawa {\it et al.} [FlowQCD Collaboration],
  Phys.\ Rev.\ D {\bf 90}, no. 1, 011501 (2014)
  Erratum: [Phys.\ Rev.\ D {\bf 92}, no. 5, 059902 (2015)]
  doi:10.1103/PhysRevD.90.011501, 10.1103/PhysRevD.92.059902
  [arXiv:1312.7492 [hep-lat]].

\bibitem{Taniguchi:2016ofw} 
  Y.~Taniguchi, S.~Ejiri, R.~Iwami, K.~Kanaya, M.~Kitazawa, H.~Suzuki, T.~Umeda and N.~Wakabayashi,
  Phys.\ Rev.\ D {\bf 96}, no. 1, 014509 (2017)
  doi:10.1103/PhysRevD.96.014509
  [arXiv:1609.01417 [hep-lat]].

\bibitem{Kitazawa:2016dsl} 
  M.~Kitazawa, T.~Iritani, M.~Asakawa, T.~Hatsuda and H.~Suzuki,
  Phys.\ Rev.\ D {\bf 94}, no. 11, 114512 (2016)
  doi:10.1103/PhysRevD.94.114512
  [arXiv:1610.07810 [hep-lat]].

\bibitem{Ejiri:2017wgd} 
  S.~Ejiri {\it et al.},
  PoS LATTICE {\bf 2016}, 058 (2017)
  doi:10.22323/1.256.0058
  [arXiv:1701.08570 [hep-lat]].

\bibitem{Kitazawa:2017qab} 
  M.~Kitazawa, T.~Iritani, M.~Asakawa and T.~Hatsuda,
  Phys.\ Rev.\ D {\bf 96}, no. 11, 111502 (2017)
  doi:10.1103/PhysRevD.96.111502
  [arXiv:1708.01415 [hep-lat]].

\bibitem{Kanaya:2017cpp} 
  K.~Kanaya {\it et al.} [WHOT-QCD Collaboration],
  EPJ Web Conf.\  {\bf 175}, 07023 (2018)
  doi:10.1051/epjconf/201817507023
  [arXiv:1710.10015 [hep-lat]].

\bibitem{Taniguchi:2017ibr} 
  Y.~Taniguchi {\it et al.} [WHOT-QCD Collaboration],
  EPJ Web Conf.\  {\bf 175}, 07013 (2018)
  doi:10.1051/epjconf/201817507013
  [arXiv:1711.02262 [hep-lat]].

\bibitem{Yanagihara:2018qqg} 
  R.~Yanagihara, T.~Iritani, M.~Kitazawa, M.~Asakawa and T.~Hatsuda,
  arXiv:1803.05656 [hep-lat].

\bibitem{Hirakida:2018uoy} 
  T.~Hirakida, E.~Itou and H.~Kouno,
  arXiv:1805.07106 [hep-lat].

\bibitem{Morikawa:2018fek} 
  O.~Morikawa and H.~Suzuki,
  PTEP {\bf 2018}, no. 7, 073B02 (2018)
  doi:10.1093/ptep/pty073
  [arXiv:1803.04132 [hep-th]].

\bibitem{Endo:2015iea} 
  T.~Endo, K.~Hieda, D.~Miura and H.~Suzuki,
  PTEP {\bf 2015}, no. 5, 053B03 (2015)
  doi:10.1093/ptep/ptv058
  [arXiv:1502.01809 [hep-lat]].

\bibitem{Hieda:2016lly} 
  K.~Hieda and H.~Suzuki,
  Mod.\ Phys.\ Lett.\ A {\bf 31}, no. 38, 1650214 (2016)
  doi:10.1142/S021773231650214X
  [arXiv:1606.04193 [hep-lat]].

\bibitem{Bochicchio:1985xa} 
  M.~Bochicchio, L.~Maiani, G.~Martinelli, G.~C.~Rossi and M.~Testa,
  Nucl.\ Phys.\ B {\bf 262}, 331 (1985).
  doi:10.1016/0550-3213(85)90290-1

\bibitem{Taniguchi:2016tjc} 
  Y.~Taniguchi, K.~Kanaya, H.~Suzuki and T.~Umeda,
  Phys.\ Rev.\ D {\bf 95}, no. 5, 054502 (2017)
  doi:10.1103/PhysRevD.95.054502
  [arXiv:1611.02411 [hep-lat]].

\bibitem{Ferrara:1974pu} 
  S.~Ferrara and B.~Zumino,
  Nucl.\ Phys.\ B {\bf 79}, 413 (1974).
  doi:10.1016/0550-3213(74)90559-8

\bibitem{Fayet:1975yi} 
  P.~Fayet,
  Nucl.\ Phys.\ B {\bf 113}, 135 (1976).
  doi:10.1016/0550-3213(76)90458-2

\bibitem{Brink:1976bc} 
  L.~Brink, J.~H.~Schwarz and J.~Scherk,
  Nucl.\ Phys.\ B {\bf 121}, 77 (1977).
  doi:10.1016/0550-3213(77)90328-5

\bibitem{Seiberg:1994rs} 
  N.~Seiberg and E.~Witten,
  Nucl.\ Phys.\ B {\bf 426}, 19 (1994)
  Erratum: [Nucl.\ Phys.\ B {\bf 430}, 485 (1994)]
  doi:10.1016/0550-3213(94)90124-4, 10.1016/0550-3213(94)00449-8
  [hep-th/9407087].

\bibitem{Sugino:2003yb} 
  F.~Sugino,
  JHEP {\bf 0401}, 015 (2004)
  doi:10.1088/1126-6708/2004/01/015
  [hep-lat/0311021].

\bibitem{Sugino:2004uv} 
  F.~Sugino,
  JHEP {\bf 0501}, 016 (2005)
  doi:10.1088/1126-6708/2005/01/016
  [hep-lat/0410035].

\bibitem{Damgaard:2007be} 
  P.~H.~Damgaard and S.~Matsuura,
  JHEP {\bf 0707}, 051 (2007)
  doi:10.1088/1126-6708/2007/07/051
  [arXiv:0704.2696 [hep-lat]].

\bibitem{Damgaard:2007xi} 
  P.~H.~Damgaard and S.~Matsuura,
  JHEP {\bf 0708}, 087 (2007)
  doi:10.1088/1126-6708/2007/08/087
  [arXiv:0706.3007 [hep-lat]].

\bibitem{Hanada:2011qx} 
  M.~Hanada, S.~Matsuura and F.~Sugino,
  Nucl.\ Phys.\ B {\bf 857}, 335 (2012)
  doi:10.1016/j.nuclphysb.2011.12.014
  [arXiv:1109.6807 [hep-lat]].

\bibitem{Takimi:2012zw} 
  T.~Takimi,
  JHEP {\bf 1208}, 069 (2012)
  doi:10.1007/JHEP08(2012)069
  [arXiv:1205.7038 [hep-lat]].

\bibitem{Wess:1992cp} 
  J.~Wess and J.~Bagger,
  Princeton, USA: Univ. Pr. (1992) 259 p

\bibitem{Capri:2018lru} 
  M.~A.~L.~Capri, S.~P.~Sorella, R.~C.~Terin and H.~C.~Toledo,
  arXiv:1801.09221 [hep-th].

\bibitem{Capponi:2015ucc} 
  F.~Capponi, A.~Rago, L.~Del Debbio, S.~Ehret and R.~Pellegrini,
  PoS LATTICE {\bf 2015}, 306 (2016)
  doi:10.22323/1.251.0306
  [arXiv:1512.02851 [hep-lat]].

\bibitem{Fujikawa:2016qis} 
  K.~Fujikawa,
  JHEP {\bf 1603}, 021 (2016)
  doi:10.1007/JHEP03(2016)021
  [arXiv:1601.01578 [hep-lat]].

\bibitem{Aoki:2016ohw} 
  S.~Aoki, J.~Balog, T.~Onogi and P.~Weisz,
  PTEP {\bf 2016}, no. 8, 083B04 (2016)
  doi:10.1093/ptep/ptw106
  [arXiv:1605.02413 [hep-th]].

\bibitem{Makino:2018rys} 
  H.~Makino, O.~Morikawa and H.~Suzuki,
  PTEP {\bf 2018}, no. 5, 053B02 (2018)
  doi:10.1093/ptep/pty050
  [arXiv:1802.07897 [hep-th]].

\bibitem{Kikuchi:2014rla} 
  K.~Kikuchi and T.~Onogi,
  JHEP {\bf 1411}, 094 (2014)
  doi:10.1007/JHEP11(2014)094
  [arXiv:1408.2185 [hep-th]].

\bibitem{Aoki:2017iwi} 
  S.~Aoki, K.~Kikuchi and T.~Onogi,
  JHEP {\bf 1802}, 128 (2018)
  doi:10.1007/JHEP02(2018)128
  [arXiv:1704.03717 [hep-th]].

\bibitem{Kadoh:2018}
D.~Kadoh and N.~Ukita,
``SYM flow equation in $N=1$ SUSY'',
talk presented at The 36th Annual International Symposium on Lattice Field Theory (LATTICE 2018),\\
\url{https://indico.fnal.gov/event/15949/session/16/contribution/106}

\bibitem{Suzuki:2015bqa} 
  H.~Suzuki,
  PTEP {\bf 2015}, no. 10, 103B03 (2015)
  doi:10.1093/ptep/ptv139
  [arXiv:1507.02360 [hep-lat]].

\bibitem{Tarasov:1980au} 
  O.~V.~Tarasov, A.~A.~Vladimirov and A.~Y.~Zharkov,
  Phys.\ Lett.\  {\bf 93B}, 429 (1980).
  doi:10.1016/0370-2693(80)90358-5

\bibitem{Avdeev:1981ew} 
  L.~V.~Avdeev and O.~V.~Tarasov,
  Phys.\ Lett.\  {\bf 112B}, 356 (1982).
  doi:10.1016/0370-2693(82)91068-1

\bibitem{Grisaru:1982zh} 
  M.~T.~Grisaru and W.~Siegel,
  Nucl.\ Phys.\ B {\bf 201}, 292 (1982)
  Erratum: [Nucl.\ Phys.\ B {\bf 206}, 496 (1982)].
  doi:10.1016/0550-3213(82)90433-3, 10.1016/0550-3213(82)90282-6

\bibitem{Gates:1983nr} 
  S.~J.~Gates, M.~T.~Grisaru, M.~Ro\v cek and W.~Siegel,
  Front.\ Phys.\  {\bf 58}, 1 (1983)
  [hep-th/0108200].

\bibitem{Seiberg:1988ur} 
  N.~Seiberg,
  Phys.\ Lett.\ B {\bf 206}, 75 (1988).
  doi:10.1016/0370-2693(88)91265-8

\end{thebibliography}
\end{document}